\documentclass[twocolumn]{aastex62}
\usepackage{
  afterpage,     
  tabularx,
  booktabs,      
  multirow,      
  physics
}
\usepackage[caption=false, position=top]{subfig}

\usepackage{savesym}
\savesymbol{tablenum}
\usepackage[
  separate-uncertainty=true,
  multi-part-units=single, 
  bracket-numbers=false,
  angle-symbol-over-decimal=true]{siunitx}
\restoresymbol{SIX}{tablenum}

\newcommand{\wes}{Department of Astronomy, Van Vleck Observatory, Wesleyan University, 96 Foss Hill Drive, Middletown, CT 06459, USA}

\begin{document}

\title{The Mass of Stirring Bodies in the AU Mic Debris Disk Inferred from Resolved Vertical Structure}
\shorttitle{Vertical Structure in the AU Mic Debris Disk}

\author{Cail Daley}
\affiliation{\wes}
\affiliation{Department of Astronomy, University of Illinois Urbana-Champaign, Urbana, IL 61801, USA}

\author{A. Meredith Hughes}
\affiliation{\wes}

\author{Evan S. Carter}
\affiliation{\wes}

\author{Kevin Flaherty}
\affiliation{\wes}
\affiliation{Department of Astronomy and Department of Physics, Williams College, Williamstown, MA 01267, USA}

\author{Zachary Lambros}
\affiliation{\wes}

\author{Margaret Pan}
\affiliation{Department of Earth, Atmospheric and Planetary Sciences, Massachusetts Institute of Technology, Cambridge, MA 02139, USA}

\author{Hilke Schlichting}
\affiliation{Department of Earth, Planetary and Space Sciences, University of California, Los Angeles, CA 90095, USA}
\affiliation{Department of Earth, Atmospheric and Planetary Sciences, Massachusetts Institute of Technology, Cambridge, MA 02139, USA}

\author{Eugene Chiang}
\affiliation{Department of Astronomy, University of California at Berkeley, Campbell Hall, Berkeley, CA 94720-3411}
\affiliation{Department of Earth and Planetary Science, University of California at Berkeley, McCone Hall, Berkeley, CA 94720-4767}

\author{Mark Wyatt}
\affiliation{Institute of Astronomy, University of Cambridge, Madingley Road, Cambridge CB3 0HA, United Kingdom}

\author{David Wilner}
\affiliation{Harvard-Smithsonian Center for Astrophysics, 60 Garden Street, Cambridge, MA, 02138, USA}

\author{Sean Andrews}
\affiliation{Harvard-Smithsonian Center for Astrophysics, 60 Garden Street, Cambridge, MA, 02138, USA}

\author{John Carpenter}
\affiliation{Joint ALMA Observatory (JAO), Alonso de Cordova 3107 Vitacura-—Santiago de Chile, Chile}

\received{August 16, 2018}
\revised{March 10, 2019}
\accepted{March 15, 2019}
\submitjournal{ApJ}

\begin{abstract}
The vertical distribution of dust in debris disks is sensitive to the number and size of large planetesimals dynamically stirring the disk, and is therefore well-suited for constraining the prevalence of otherwise unobservable Uranus and Neptune analogs.
Information regarding stirring bodies has previously been inferred from infrared and optical observations of debris disk vertical structure, but theoretical works predict that the small particles traced by short-wavelength observations will be `puffed up' by radiation pressure, yielding only upper limits. 
The large grains that dominate the disk emission at millimeter wavelengths are much less sensitive to the effects of stellar radiation or stellar winds, and therefore trace the underlying mass distribution more directly. 
Here we present ALMA \SI{1.3}{mm} dust continuum observations of the debris disk around the nearby M star AU Mic.  
The \SI{3}{au} spatial resolution of the observations, combined with the favorable edge-on geometry of the system, allows us to measure the vertical thickness of the disk.
We report a scale height-to-radius aspect ratio of $h = 0.031_{-0.004}^{+0.005}$ between radii of $\sim \SI{23}{au}$ and $\sim \SI{41}{au}$. 
Comparing this aspect ratio to a theoretical model of size-dependent velocity distributions in the collisional cascade, we find that the perturbing bodies embedded in the local disk  must be larger than about \SI{400}{km}, and the largest perturbing body must be smaller than roughly $\SI{1.8}{M_\earth}$.
These measurements rule out the presence of a gas giant or Neptune analog near the $\sim \SI{40}{au}$ outer edge of the debris ring, but are suggestive of large planetesimals or an Earth-sized planet stirring the dust distribution.
\end{abstract}

\keywords{Stars: circumstellar matter, Stars: individual (AU Mic), Planetary Systems: planet--disk interactions, Submillimeter: planetary systems}

\section{Introduction}
\label{section: introduction}

Planets form during a relatively short and early stage in the lifetime of stellar systems,  when the host star is still encircled by a protoplanetary disk.
Planet formation, as well as processes including accretion, photoevaporation, and winds, causes first-generation protoplanetary material to dissipate over time \citep{williams&cieza11,ercolano&pascucci17}.
The first-generation material is replaced by second-generation `debris,' produced by collisional grinding of larger planetesimals into small dust grains in a process known as a collisional cascade \citep{wyatt2008}. 
The resulting debris disks, optically thin and significantly less luminous than their protoplanetary counterparts, are currently observable around at least 25\% of Solar-type stars and are likely to be at least as common as the exoplanetary systems with which they are thought to be associated \citep{montesinos16}.

Analysis of the morphological and emissive properties of debris disks sheds light on the final stages of planetary system evolution and can reveal the presence of planets hidden within.
Planets can imprint features such as rings, gaps, clumps, or other asymmetries on their parent disks, although it is rarely straightforward to infer the properties of planets directly from the disk morphology \citep[see the review by][and references therein]{hughes18}.
In gas-poor systems, the vertical structure of a debris disk can serve as a probe of the total mass of large bodies stirring the collisional cascade \citep{thebault09}.
The presence of massive bodies increases the inclination dispersion of the dust particle orbits and thus the scale height $H$ of the observed dust distribution.

The dynamical excitation of a disk can therefore be determined from its aspect ratio $H/r$, which in turn allows inferences about the mass and size of bodies responsible for the dynamical stirring.
Such work has been undertaken by several authors using visible and infrared observations \citep{artymowicz97,thebault&augereau07,quillen07}.
However, \citet{thebault09} demonstrates that radiation pressure from the host star should preferentially excite the smallest dust grains in a disk, imparting a `natural' scale height to the system even in the absence of large stirring bodies. 
Thus longer-wavelength ($\lambda \geq \SI{50}{\mu m}$ for typical grain blow-out sizes of $\sim 2$ to \SI{10}{\mu m}) observations are required to measure the disk scale height resulting from dynamical stirring alone, as the large grains dominating the emission at these wavelengths are much less sensitive to the effects of radiation pressure.

The M3IVe star AU Mic presents a particularly favorable target for such observations because of its proximity (\SI{9.725 \pm 0.005}{pc}; \citealp{gaia16,gaia_DR2}), edge-on inclination, and apparently symmetric morphology at millimeter wavelengths.
The first M star detected to have a far-infrared excess, AU Mic hosts one of the best-studied debris disks \citep{moshir90}. 
As a member of the $\beta$ Pic Moving Group, it is thought to be relatively young: $\SI{23 \pm 2}{Myr}$ \citep{binks&jeffries14,mamajek&bell14,malo14}. 
A wide range of stellar masses \citep[\SIrange{0.3}{0.6}{M_\sun};][]{plavchan09,houdebine&doyle94} are reported in the literature; \citet{schuppler15} assume a stellar mass of \SI{0.5}{M_\sun} based on the mean of these values.
The disk around AU Mic was first resolved by \citet{kalas04} in scattered light, and a host of observations spanning the optical to the submillimeter have followed \citep{augereau&beust06,macgregor13,schneider14,matthews15,wang15}. 

Notably, the radial structure of the debris around AU Mic exhibits a so-far-unique time variability at scattered light wavelengths.
\citet{boccaletti15,boccaletti18} identify several local intensity maxima offset from the disk midplane.
{On the southeast, these features are moving away from the star} at projected velocities that are not consistent with Keplerian rotation; in fact, the outermost features appear to be unbound from the star. 
\citet{sezestre17} provide kinematic fits and invoke a dust source of unspecified nature to explain the features.
\citet{chiang&fung17} propose that these fast-moving features are made up of dust particles repelled by a time-variable stellar wind that triggers dust avalanches when the wind blows strongest. 
These avalanches would be seeded by the debris remaining from the recent disruption of a $\sim \SI{400}{km}$ sized progenitor. 
In this paper we will provide an independent constraint on the presence of comparably sized planetesimals.

We present here \ang{;;0.3} Atacama Large Millimeter/sub\-millimeter Array (ALMA) \SI{1.3}{mm} observations of the AU Mic debris disk. 
These observations represent a factor of $\sim 2$ improvement in both spatial resolution and rms noise relative to previous ALMA observations of the system by \citet{macgregor13}, and our analysis indicates that the vertical structure of the disk is resolved with $4 \sigma$ confidence.
In \S \ref{section: observations} we present the new observations and describe the data reduction.  
In \S \ref{section: results} we document the basic observational results regarding the disk flux, morphology, and gas content.  
In \S \ref{section: analysis} we conduct a parametric exploration of an axisymmetric disk model in order to investigate the degeneracy between vertical structure, radial structure, and viewing geometry.
In \S \ref{section: discussion} we discuss our results, particularly the constraints on the dynamical excitation of the disk imposed by our measurement of the scale height, and compare them to previous observations.
In \S \ref{section: conclusion} we summarize the results of our scale height measurement and its implications for the population of stirring bodies in AU Mic's disk.

\section{Observations}

\begin{table*}[t]
  \centering
  \begin{tabular*}{\textwidth}{l @{\extracolsep{\fill}} rrrr}
  \toprule

  {Observational parameters}
                            & 2014 Mar 26    & 2014 Aug 18    & 2015 Jun 24   \\
  \cmidrule(lr){2-4} 
  \cmidrule(lr){1-1} 
  Stellar RA (J2000):       & 20:45:09.8424  & 20:45:09.8543  & 20:45:09.8719 \\
  Stellar DEC (J2000):      & -31.20.32.360  & -31.20.32.522  & -31.20.32.839 \\
  Antennas:                 & 32             & 35             & 37            \\
  Baseline length (m):      & 12--406        & 19--1160       & 30--1320      \\
  On-source time (min):     & 35             & 35             & 33            \\
  Flux calibrator:          & Titan          & J2056-472      & Titan         \\
  Bandpass calibrator:      & J1924-2914     & J2056-4714     & J1924-2914    \\
  Gain calibrator:          & J2101-2933     & J2101-2933     & J2056-3208    \\
  pwv range (mm):           &[0.63, 0.66]    & [1.58, 1.69]   & [0.67, 0.74]
  \vspace{1em}                                                                \\

  {Imaging parameters} &&&                                                    \\
  \cmidrule(lr){1-1}
  Beam size (arcsec): & 
    $1.27\times0.74$ & 
    $0.33\times0.30$ & 
    $0.47\times0.31$                                                          \\
  Peak intensity (\si{\mu Jy.beam^{-1}}): & 630 & 240 & 320                      \\
  rms noise (\si{\mu Jy.beam^{-1}}):      &  30 &  30 &  20                      \\
  \bottomrule
  \end{tabular*}

	\caption{
  Observational and imaging parameters for the three datasets used in this work. 
  Images were created using the task \texttt{tclean} with natural weighting.}
  \label{tab:observations}
\end{table*}

\label{section: observations}
AU Mic was observed with ALMA on three separate occasions in 2014 March, 2014 August and 2015 June (see Table \ref{tab:observations}).
All observations employed ALMA's 12m antennas and Band 6 receivers, including four independently tunable spectral windows each with a bandwith of \SI{1.875}{GHz}. 
One spectral window was centered around the CO~$\mathrm{J}=2-1$ transition at a rest frequency of 230.538001 GHz and a channel spacing of \SI{488}{\kHz} (\SI{0.635}{km/s}).
The remaining three spectral windows were configured to detect continuum emission with central frequencies of 228.5, 213.5, and \SI{216.0}{GHz} and channel spacings of \SI{15.625}{Mhz} ($\sim$\SI{21}{km/s}).
The mean of the four central wavelengths is \SI{1.35}{mm}.
The baseline lengths range between \SIlist{12;1320}{m}; the longest baseline among the three observations traces an angular scale of \ang{;;0.22} and a spatial scale of \SI{2.1}{au}.

Calibration, reduction, and imaging were carried out using the \texttt{CASA} software package. 
Standard ALMA reduction scripts were applied to the datasets: phase calibration was accomplished via gain calibration and water vapor radiometry tables, while system temperature calibrations were performed to account for variations in instrument and weather conditions. 
Flux and bandpass calibrations were subsequently applied; the flux calibration is subject to a 10\% systematic uncertainty.
Both \SI{2}{GHz} spectral averaging and \SI{60}{s} spectral averaging were performed.
In addition to these standard procedures, the weights of the visibilities were recalculated using the variance around each baseline as in \citet{flaherty17}.

During the last segment of the June observation (04:23:38-04:29:58 UT), the host star flared. 
To determine the flux density of the flare as a function of time, we first binned the data into one-minute intervals using the task \texttt{split}. 
Fitting a point source to the long baselines in each bin tended to overestimate the stellar flux when compared with the results of our parametric modeling presented in \S \ref{section: analysis}, as even the longest baselines are sensitive to disk emission.
Instead, we used the task \texttt{tclean} to image the emission with natural weighting and determined the flux density of the peak pixel in each image. 
To account for contamination of the stellar component by disk emission, we also subtracted the disk-only flux density in the corresponding pixel of our best-fit disk model image convolved with the ALMA visibilities.
The resulting flare flux densities can be found in Table \ref{tab:flare fluxes}; details of the model fitting process and a list of best-fit parameters can be found in \S \ref{section: analysis} and Table \ref{tab: params}, respectively.
We exclude from our analysis of the disk emission the six minutes during which the flare occurred, as it proved difficult to separate the stellar emission from that of the disk while it was changing so rapidly.
The peak flux density in the no-flare segment of the June observations is consistent with the best-fit model stellar flux within uncertainties (Table \ref{tab: params}), suggesting that this method suffers minimal contamination from disk emission.

\begin{table}[b]
  \centering
  \begin{tabular}{lr}
    \toprule
    Time (UTC) & Peak $F_\nu$ (\si{mJy.beam^{-1}}) \\
    \midrule
    03:45:0--04:20:0 (no flare) & $0.26 \pm 0.02$\\
  	4:23:38--4:24:00 & $0.84  \pm 0.15$ \\
  	4:24:00--4:25:00 & $11.10 \pm 0.12$ \\
  	4:25:00--4:26:00 & $3.46  \pm 0.10$ \\
  	4:26:00--4:27:00 & $1.40  \pm 0.10$ \\
  	4:27:00--4:28:00 & $0.48  \pm 0.11$ \\
  	4:28:00--4:29:00 & $0.40  \pm 0.10$ \\
  	4:29:00--4:29:58 & $0.33  \pm 0.10$\\
    \bottomrule
  \end{tabular}
	\caption{The peak flux densities of the central source before and during the June 24 flare, determined from naturally-weighted clean images of one-minute time bins.
        Uncertainties are given by the off-source rms noise of each image.}
  \label{tab:flare fluxes}
\end{table}

Imaging was performed using standard Fourier inversion methods as implemented in the \texttt{CASA} task \texttt{tclean}. 
Visual inspection of the cleaned images for each of the three dates indicated that the peak of the bright chromospheric emission from the central star was consistently offset from the image center by roughly \ang{;;0.3} in each direction.
To obtain a more precise alignment for each of the datasets, we fit an image-domain 2-D Gaussian to a small region around the star with the task \texttt{imfit}, and used the centroid of the Gaussian fit to define the star position.
Each dataset was then phase-shifted using the task \texttt{fixvis} so that the pointing center of the data was the same as the fitted star position. 
We note that the centroid of the Gaussian fit to the flare segment of the June observations is \SI{1.13 \pm 0.18}{au} (30\% of the synthesized beam) NE of the non-flare fit centroid.  
This change in the fit stellar location might be explained if the flare were not symmetric with respect to the star, although such a large offset seems unlikely as it is orders of magnitude larger than the stellar radius; an issue with the position of one of the calibrators may also be responsible.

AU Mic is a high proper motion system, and its equatorial coordinates changed significantly over the 1.5 years between the first and last observations.  
Because the \texttt{CASA} task \texttt{tclean} preserves pointing center offsets when converting several visibility datasets into an image, it was necessary to combine the data into a single measurement set before creating a composite image in order to account for the offset in phase center between datasets. 
This was done using the task \texttt{concat}, which combines datasets with pointing centers aligned so long as their pointing centers do not differ by a value greater than the parameter \texttt{dirtol}.
A natural weighting scheme was used to trace small-scale disk structure, resulting in an rms noise of \SI{15}{\mu Jy.beam^{-1}} and a $\ang{;;0.52} \times \ang{;;0.39}$ restoring beam with a position angle (PA) of \ang{77.9}. 

\begin{figure}[t]
    \centering
    \includegraphics[width=\linewidth]{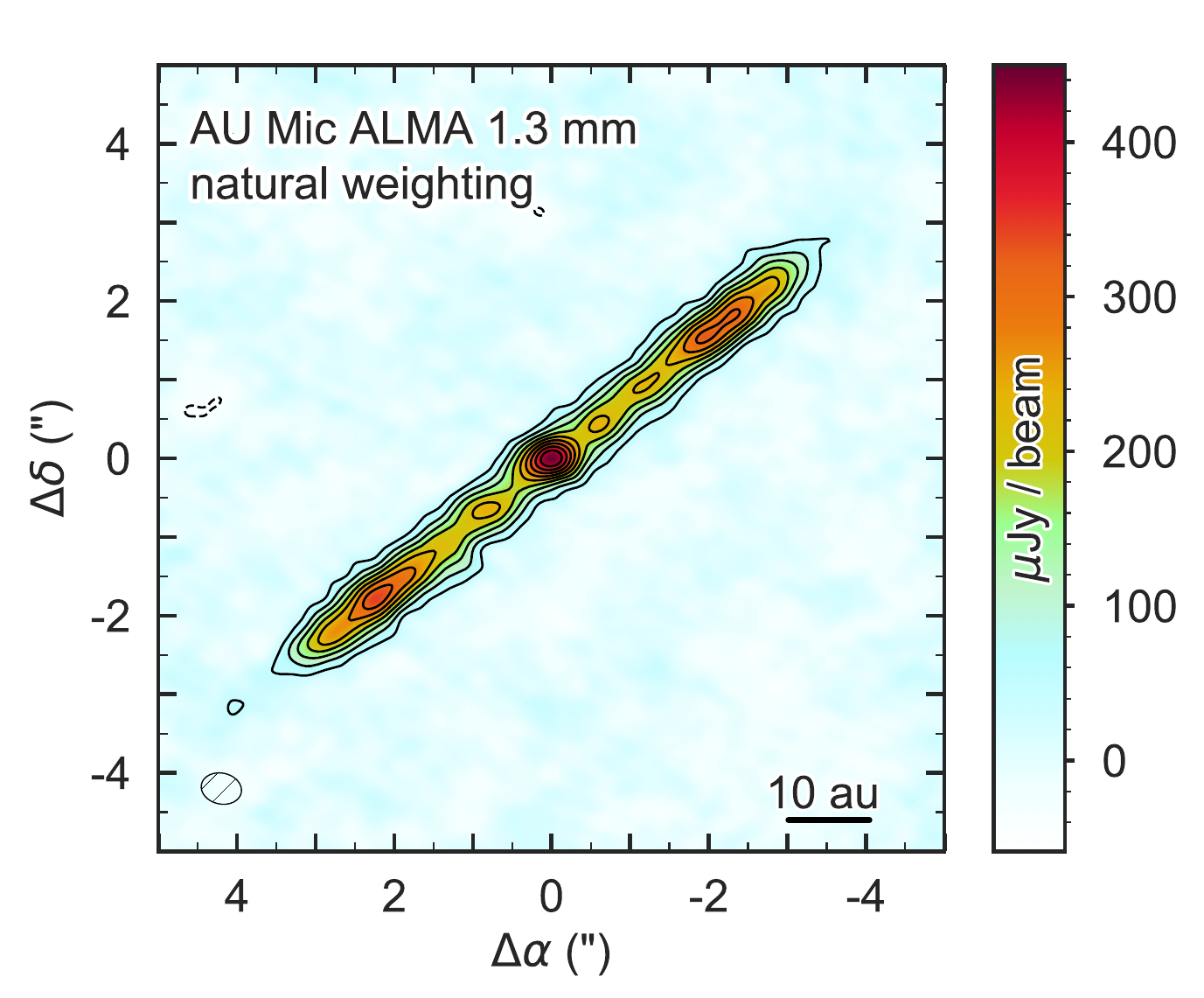}
    \caption{The AU Mic system imaged by ALMA at a wavelength of \SI{1.3}{mm} using natural weighting. 
    The rms noise is \SI{15}{\mu Jy.beam^{-1}}, and the restoring beam has dimensions $\ang{;;0.52} \times \ang{;;0.39}$ with a PA of \ang{77.9}.
    Contours are integer multiples of the three times the rms noise.
    The hatched ellipse in the bottom left of each pane designates the size and shape of the restoring beam.
    }
    \label{fig: aumic_imaged}
\end{figure}

\section{Results}
\label{section: results}

Figure \ref{fig: aumic_imaged} shows the combined dust continuum emission from all three observations at \SI{1.3}{mm}; chromospheric emission from the M star is visible as a point source at the center of the image \citep{cranmer13}. 
The peak signal-to-noise ratio of the dust emission is $\sim 23$.
Using the \texttt{MIRIAD} task \texttt{cgcurs}, we measure an integrated flux density of \SI{4.97 \pm 0.08}{\milli Jy} enclosed within the $3\sigma$ contours of the naturally weighted image.  
We note that this value represents the combined emission from the disk \textit{and} the star. 
Faithfully disentangling the two components proved difficult, both because the stellar flux varied significantly across the three nights of observation (Table \ref{tab: params}) and because emission from the star and edge-on disk overlap in the sky plane.
Consequently, the most accurate way to isolate the disk flux from the stellar contribution is through parametric modeling (see \S \ref{section: analysis}) where the two components can be specified separately; our modeling yields a disk flux density of \SI[parse-numbers=false]{4.81 ^{+0.04} _{-0.05}}{mJy}.
Our estimate of the disk flux density is significantly smaller than the best-fit \SI{1.28}{mm} disk flux density of \SI[parse-numbers=false]{7.14^{+0.12}_{-0.25}}{mJy} reported by \citet{macgregor13} (equivalent to \SI[parse-numbers=false]{6.42^{+0.11}_{-0.22}}{mJy} when scaled to our observing wavelength of \SI{1.35}{mm} assuming a millimeter-wavelength spectral index of 2 \citep{matthews15}.
The $\sim 25\%$ discrepancy is unexpectedly large, but may fall within the combined absolute flux calibration uncertainties of the two datasets.
Because our shortest baseline ($\sim \SI{9}{k\lambda}$) is smaller than the shortest baseline in \citet{macgregor13} ($\sim \SI{16}{k\lambda}$), we cannot attribute the flux discrepancy to extended emission that might have been unresolved by our observations.

The ansa to the SE exhibits a maximum flux density of \SI{340 \pm 15}{\mu Jy.beam^{-1}} at a stellocentric separation of \SI{29.0 \pm 0.2}{au} and PA of $\ang {128.7} \pm 0.4$, while the ansa to the NW exhibits a maximum flux density of \SI{330 \pm 15}{\mu Jy.beam^{-1}} at a separation of \SI{24.1 \pm 0.2}{au} and PA of $\ang{308.2} \pm 0.4$.
Here PA is measured counterclockwise with respect to the north celestial pole.
The discrepancy in position angle between the two peaks falls within the estimated uncertainties and so we are not able to confirm the scattered light PA offset observed by \citet{boccaletti15}. 
The peak flux densities of the ansae differ by less than the rms noise, indicating that there is no significant difference in brightness between the two sides of the disk.
Indeed, the apparent flux asymmetry in these data is in the opposite direction of the apparent flux asymmetry in \citet{macgregor13}, providing further circumstantial evidence that there is no significant flux asymmetry at millimeter wavelengths. 

\begin{figure}[p]
    \centering
    \includegraphics[width=\linewidth]{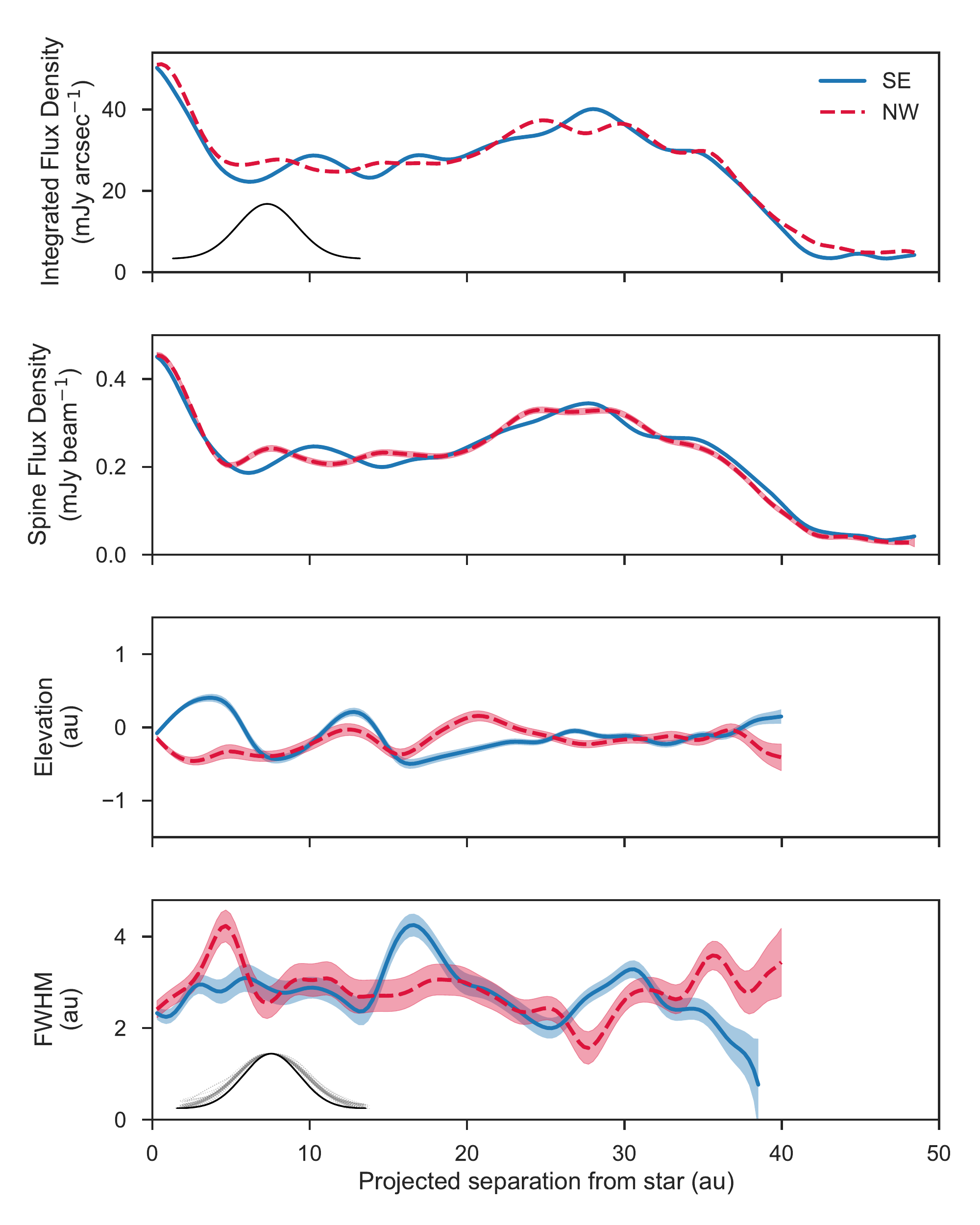}
    \caption{The AU Mic debris disk's radial and vertical structure, extracted from the naturally weighted image in Figure \ref{fig: aumic_imaged}. 
    The solid blue line shows the southeast limb of the disk, while the dashed red line shows the northeast limb; the shaded bands designate $1\sigma$ uncertainties.
    From top to bottom the four panes show, as a function of projected separation from the star, i) the vertically integrated surface brightness, ii) the disk spine surface brightness, iii) the disk spine deviation from the midplane, and iv) the beam-deconvolved disk FWHM.
    Because the disk is viewed edge-on and a wide range of stellocentric separations contribute to the FWHM at a given projected separation, the apparent vertical FHWM has no radial dependence.
    The Gaussian profile inset in the first pane shows the size of the combined dataset synthesized beam projected onto the \textit{radial} axis of the disk, while the profile inset in the bottom pane shows its projection onto the \textit{vertical} axis.
    The bottom inset also shows 40 disk vertical profiles (dotted lines) sampled at random from $r \leq \SI{40}{au}$; the profiles have all been scaled to the same amplitude and centroid, showing that the FWHMs of the disk profiles are larger than that of the beam.
    Although the last pane indicates a beam-subtracted disk FWHM that is technically smaller than the vertical FWHM of the beam, the fact that the image-domain vertical height of the disk is consistently in excess of the beam contribution implies that our data spatially resolve the vertical structure of the disk.
    }
    \label{fig: boccaletti}
\end{figure}

While comparison of the ansae peak flux densities suggests that the AU Mic disk does not exhibit severe asymmetry, a more detailed analysis can be conducted by extracting surface brightness and vertical structure profiles from the naturally weighted image in Figure \ref{fig: aumic_imaged}.
These profiles are shown in Figure \ref{fig: boccaletti}.
The top pane was created by integrating the disk vertical surface brightness profile (resulting in units of \si{mJy.arcsec^{-1}} after accounting for the beam area) at a series of slices along the disk major axis, while the remaining three panes were created by fitting one-dimensional Gaussians to these slices.
The second, third, and fourth panes correspond respectively to the amplitude, centroid, and beam-subtracted FWHM of the Gaussian fits. 
Broadening effects of the synthesized beam in the vertical direction have been removed by subtracting in quadrature the Gaussian beam FWHM along the PA of the disk minor axis from the Gaussian fit FWHM.
In more physical terms, the first pane represents a model-independent surface brightness profile, while the remaining three panes assume Gaussian vertical structure and show the flux density at the location of the disk `spine,' the spine's elevation from the midplane, and the disk FWHM.

As seen in the top two panes of Figure \ref{fig: boccaletti}, the surface brightness profiles of the two ansae generally mirror each other.
However, we note the presence of slightly asymmetric local intensity maxima in the inner regions of the disk (especially clear in the spine surface brightness profile), one $\sim$\SI{10.2}{au} to the SE of the star and one $\sim$\SI{7.6}{au} to the NW, in addition to the previously mentioned asymmetry in the locations of the two ansae peaks.
It is unclear whether these are real features of the disk or artifacts of the rms noise or cleaning process; we examine the significance of these features in \S \ref{section: analysis}.
Three-sigma emission, as determined from the spine surface brightness profile, extends a radial distance of $\sim$\SI{42}{au} to the NW and  $\sim$\SI{44}{au} to the SE.
The disk is resolved across $\sim$20 beams along the major axis. 
We were unable to detect (in either the combined dataset or the three individual epochs) the intensity variations or excursions of the disk spine from the midplane that characterize the fast-moving features observed by \citet{boccaletti15,boccaletti18}.
This is not entirely surprising, as both \citet{sezestre17} and \citet{chiang&fung17} suggest that the features are composed of sub-micron-sized grains which do not emit efficiently in the millimeter.

Cursory analysis indicates the disk is marginally resolved perpendicular to the major axis as well, exhibiting a vertical FWHM of $\SI{2.8}{au} \approx 2/3$ beam after taking into account the broadening effects of the beam (Figure \ref{fig: boccaletti}, bottom pane).
Our ability to resolve the vertical structure of the AU Mic debris disk is discussed in greater detail in \S \ref{subsection: vertical discussion}.

\subsection{CO Content}
\label{subsection: gas}

There is no evidence that the AU Mic system harbors a significant reservoir of molecular gas.
We set a $3 \sigma$ upper limit of \SI{0.07}{Jy.km.s^{-1}} on the CO~$\mathrm{J}=2-1$ integrated flux, obtained by integrating the flux density in an $\ang{;;8} \times \ang{;;8}$ box around the star between $V_{LSRK}$ velocities of $-\SI{10}{km.s^{-1}}$ and \SI{10}{km.s^{-1}}.
For a given excitation temperature, an upper limit on the CO gas mass can be inferred from the upper limit on integrated flux.
Assuming local thermal equilibrium (LTE) and that the gas is cospatial with the dust disk, we find a total CO gas mass upper limit between \SIrange[range-phrase=\ and\ ]{1.7e-7}{8.7e-7}{M_\earth} for excitation temperatures between \SIrange[range-phrase=\ and\ ]{10}{250}{K}.

\section{Analysis}
\label{section: analysis}
Previous studies of the scale height of debris disks have demonstrated a degeneracy between vertical structure, radial structure, and viewing geometry \citep[e.g.][]{milli14}.  
For example, it can be difficult to distinguish a disk that is vertically thin but slightly inclined from one that is vertically broad but perfectly edge-on.
In light of this, we adopt a modeling approach that combines a simple ray-tracing code to properly project the radial and vertical flux distribution of the optically thin emission onto the sky plane with an MCMC fitting algorithm that allows us to explore the degree to which these known degeneracies impact our ability to measure the vertical structure of the disk.  

\subsection{Modeling Formalism}

We use the parametric structure and ray tracing disk code described in \citet{flaherty15}, itself an adaptation of earlier work by \citet{rosenfeld13}.
Synthetic sky-projected images are generated from a given temperature and density structure and are subsequently Fourier transformed to create model visibilities that can be directly compared to the interferometric data.

We assume the disk to be azimuthally symmetric and vertically isothermal. 
At a given radius $r$, the vertical density profile is assumed to be Gaussian with a standard deviation equal to the scale height $H(r)$.
The scale height is given by $H(r) = hr$, where the aspect ratio $h$ is a constant.
It is common for debris disk models to assume that scale height is a linear function of radius \citep{sai15,oloffson16}, and the theoretical work of \citet{thebault09} on vertical structure in debris disks also assumes a `global' aspect ratio.
We cannot justify choosing a more complex parameterization of the scale height given the resolution of the data compared to the FWHM of the disk. 
A Gaussian vertical structure is consistent with \citet{brown01}, who finds that the ecliptic inclination distribution of Kuiper Belt Objects (KBOs) is well described by the sum of two Gaussians.
Furthermore, the author notes that a Gaussian appears to be a ``natural functional form'' for the distribution of ecliptic inclinations in the Kuiper Belt; Monte Carlo simulations of dynamical interactions in a disk with initial inclinations of zero produce an ecliptic inclination distribution that is perfectly fit by a Gaussian.

The dust opacity is set to \SI{2.3}{\cm^2.\gram^{-1}} \citep{beckwith90}, placing the model disk in the optically thin regime for the range of dust masses explored.
For an optically thin disk, the observed thermal emission is determined by both the surface density and temperature of the dust; to break the degeneracy between these two parameters, we assume that the dust grains are in blackbody equilibrium with the central star.
{While true dust grain temperatures have been shown to deviate from temperatures given by the blackbody assumption \citep{pawellek14,pawallek15}, these deviations are negligible for grain sizes larger than $\sim \SI{100}{\mu m}$ and are not expected to affect temperature estimates for the $\sim \SI{1}{mm}$ grains traced by our ALMA observations.}
Thus the dust temperature at a distance $r$ from the host star is given by
\begin{align}
  T_{dust} (r) &= \left( \frac{L_{\star}}{16 \pi r^2 \sigma} \right)^{1/4}
\end{align}
where $L_{\star}$ is the bolometric luminosity of the star and $\sigma$ is the Stefan-Boltzmann constant. We assume that the radial surface density takes the form of a power law: 
\begin{align}
  \Sigma(r) &= 
  \begin{cases}
    \Sigma_c \, r^{p} \; \; \; \; & r_{in} \leq r \leq r_{out} \\
    0 \; \; \; \; &\mbox{otherwise} 
  \end{cases}
\end{align}
where $p$ is the power law exponent, and $r_{in}$ and $r_{out}$ are the disk inner and outer radius. 
The critical surface density $\Sigma_c$ normalizes the surface density structure for a given total dust mass $M_{dust}$:
\begin{align}
\Sigma_c &= \frac{M_{dust} \left(p + 2 \right)}{2 \pi \left[ r_{out}^{(p+2)} - r_{in}^{(p+2)} \right]}.
\end{align}

\begin{figure*}
  \centering
  \includegraphics[width=\linewidth]{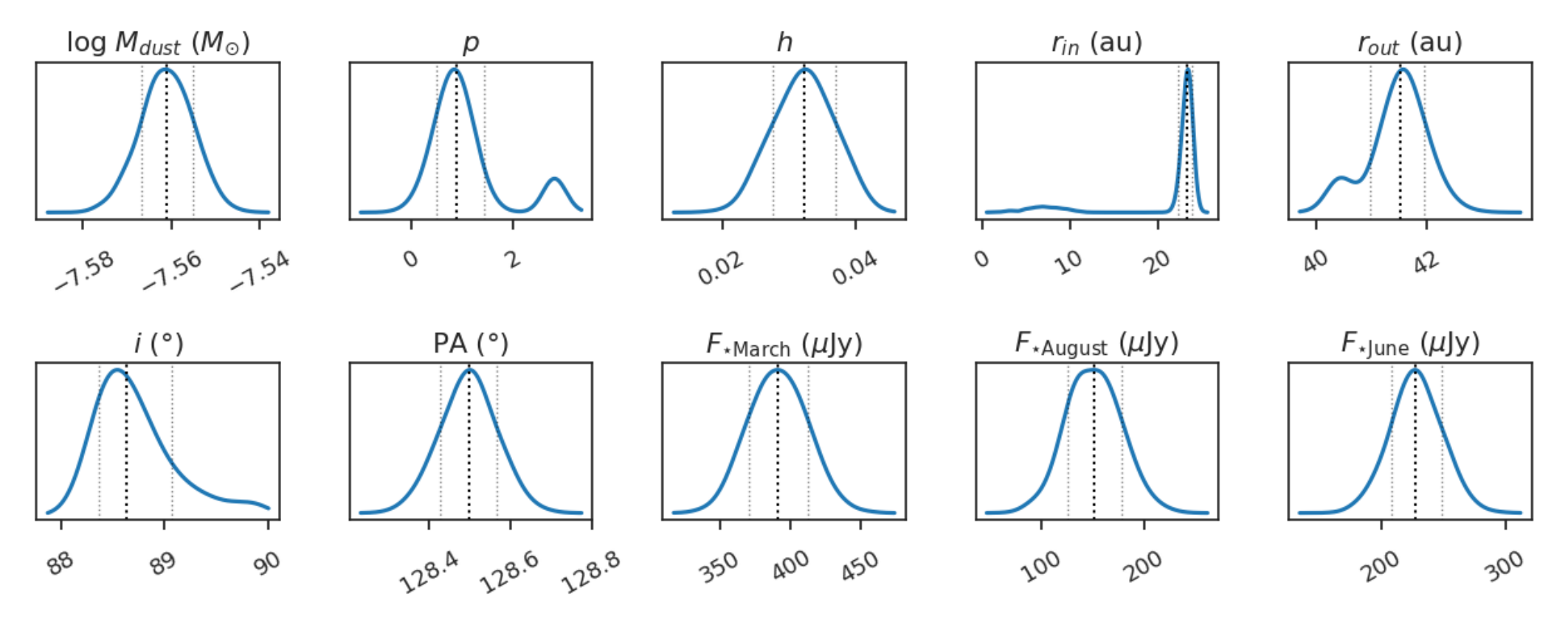}
  \caption{Kernel density estimates of the marginalized posterior probability distributions for the fiducial run. The central dashed line designates the median of each distribution while the outer lines mark the 16th and 84th percentiles ($1\sigma$ confidence intervals).}
  \label{fig: kde}
\end{figure*}

The observed disk PA and inclination $i$ are free parameters.
We adopt a stellar luminosity of \SI{0.09}{L_\sun} \citep{plavchan09} and a distance to the star of \SI{9.725 \pm 0.005}{pc}
\citep{gaia_DR2}; the observed \SI{1.3}{mm} stellar flux density $F_\star$ is left as a free parameter for each observation date.
We note that the uncertainty in the stellar distance could affect the modeled disk mass and disk radial extent, while the 10\% systematic flux uncertainty could affect the modeled disk mass and stellar flux density.
The disk center is set by the measured position of the central point source, which was obtained in \S \ref{section: observations}.
The spatial resolution of the resulting model sky image is set to \SI{0.15}{au} per pixel, $\sim \SI{7}{\percent}$ of the spatial scale sampled by the longest baseline in the data. 
After the model image is generated by the ray tracing code, it is Fourier transformed into the visibility domain and sampled at the same spatial frequencies as the ALMA data with the \texttt{MIRIAD} task \texttt{uvmodel}.
This allows the model to be compared directly to the visibilities in the Fourier domain, where uncertainties are better characterized than in the image domain.

We explore the parameter space of the model using the affine-invariant formulation of the MCMC algorithm described by \citet{goodmanweare10} and implemented in Python as \texttt{emcee} \citep{foreman-mackey13}.  
MCMC routines sample parameter space such that the density of samples in a given region is proportional to the local probability density, allowing estimation of the posterior probability functions themselves.
The process therefore not only identifies regions of high probability in parameter space, but also allows uncertainties and degeneracies between parameters to be determined from the correlations between the posteriors of each parameter. 
A log-likelihood metric $\ln \mathcal{L} = -\chi^2 / 2$ is used to assess the quality of fit between the synthetic and observed visibilities.

We assume uniform priors for all parameters.
The dust mass was sampled in logarithmic space, formally equivalent to assuming a log uniform prior.
Bounds placed on the logarithm of the dust mass were chosen to be wide enough to encompass all currently detectable optically thin dust masses, and bounds placed on the power law exponent were chosen to include all known circumstellar disk power law exponents.
Priors placed on the stellar flux density and disk inner radius, width, and aspect ratio ensured that these parameters were always greater than zero.
The position angle was confined to the range $\ang{0} < \text{PA} < \ang{180}$, and the inclination to $\ang{0} < i < \ang{90}$.
Because AU Mic is so close to edge-on, the preferred inclination falls very close to the \ang{90} prior upper bound. 
To ensure that the proximity of the solution to the edge of parameter space does not affect the posterior distribution, we investigated the effect of allowing the inclination to vary above \ang{90}. 
Doing so produced a inclination distribution with two symmetric modes on either side of \ang{90}. 
When the resulting ’unbounded’ inclination distribution was reparameterized such that all values fell between \ang{0} and \ang{90}, the original ‘bounded’ inclination distribution was recovered, indicating that the placing a \ang{90} prior upper bound has no significant effect on the inclination posterior.

\begin{table*}
  \centering
  \caption{MCMC Fitting Results}
  \label{tab: params}
  \renewcommand{\arraystretch}{1.2}
  \begin{tabular*}{\textwidth}{l @{\extracolsep{\fill}} rrrr}
  \toprule
    \multirow{2}{*}{Parameter} & \multicolumn{2}{c}{Fiducial} & \multicolumn{2}{c}{Disk + Ring} \\ 
    \cmidrule(lr){2-3} \cmidrule(lr){4-5} 
    & Median & Best Fit & Median & Best Fit \\
  \midrule
    $\log M$ (\si{M_\sun})     & $-7.555 _{-0.006} ^{+0.006}$ & $-7.557$  & $-7.548  _{-0.007} ^{+0.007}$ & $-7.545$ \\
    $p$                        & $0.9    _{-0.4}   ^{+0.5}$   & $0.9$     & $0.8     _{-0.4}   ^{+0.6}$   & $0.9$     \\
    $h$                        & $0.031  _{-0.004} ^{+0.005}$ & $0.029$   & $0.027   _{-0.005} ^{+0.004}$ & $0.028$   \\
    $r_{in}$ (\si{au})         & $23.2   _{-0.8}   ^{ +0.6}$  & $23.3$    & $24.1    _{-0.9}   ^{+0.6}$   & $23.9$    \\
    $r_{out}$(\si{au})         & $41.5   _{-0.5}   ^{ +0.4}$  & $41.5$    & $42.3    _{-0.5}   ^{+0.4}$   & $42.4$    \\
    $i$ (\si{\degree})         & $88.5   _{-0.2}   ^{ +0.3}$  & $88.5$    & $88.27   _{-0.16}  ^{+0.22}$  & $88.3$    \\
    PA  (\si{\degree})         & $128.50 _{-0.07}  ^{+0.07}$  & $128.50$  & $128.50  _{-0.07}  ^{+0.07}$  & $128.48$  \\
    March $F_*$ (\si{\mu Jy})  & $390    _{-20}    ^{+20}$    & $400$     & $390     _{-20}    ^{+20}$    & $370$     \\
    August $F_*$ (\si{\mu Jy}) & $160    _{-30}    ^{+20}$    & $170$     & $150     _{-30}    ^{+20}$    & $140$     \\
    June $F_*$ (\si{\mu Jy})   & $240    _{-20}    ^{+20}$    & $240$     & $220     _{-20}    ^{+20}$    & $210$     \\
    $r_{ring}$ (\si{au})       &                              &           & $11.9    _{-1.8}   ^{+1.7}$   & $10.8$      \\
    $M_{ring}$ ($\si{M_\earth} \times 10^{-4})$  &            &           & $1.8     _{-0.7}   ^{+0.5}$   & $1.7$     \\
    $\ln \mathcal{L}$          & $-313036 _{-4}     ^{+2}$    & $-313031$ & $-313077 _{-5}     ^{+2}$     & $-313072$ \\
  \bottomrule
\end{tabular*}
\end{table*}

We performed several MCMC runs in order to investigate a variety of model formalisms. 
All runs used 50 walkers, and $10^5$ samples were drawn from each run to allow accurate statistical comparison between runs.
Initially we varied ten parameters: the logarithm of the disk dust mass ($\log M_{dust}$), the disk inner radius ($r_{in}$), width ($\Delta r$), power law exponent ($p$), scale height aspect ratio ($h$), inclination ($i$), position angle (PA), and finally a separate stellar flux density ($F_\star$) for each of the three observation dates. 
After the fact, the posterior distribution for $\Delta r$ was replaced by the outer radius posterior $r_{out} = r_{in} + \Delta r$ to allow for easier interpretation.
This model formalism resulted in a best-fit $\chi^2$ value of {626061.8 (reduced ${\chi^2=1.032}$)}, and we treat this parameterization as our fiducial model.

\subsection{Investigating Radial Structure}
\label{subsection: radial analysis}

\begin{figure*}[p]
  \centering
  \includegraphics[width=\linewidth]{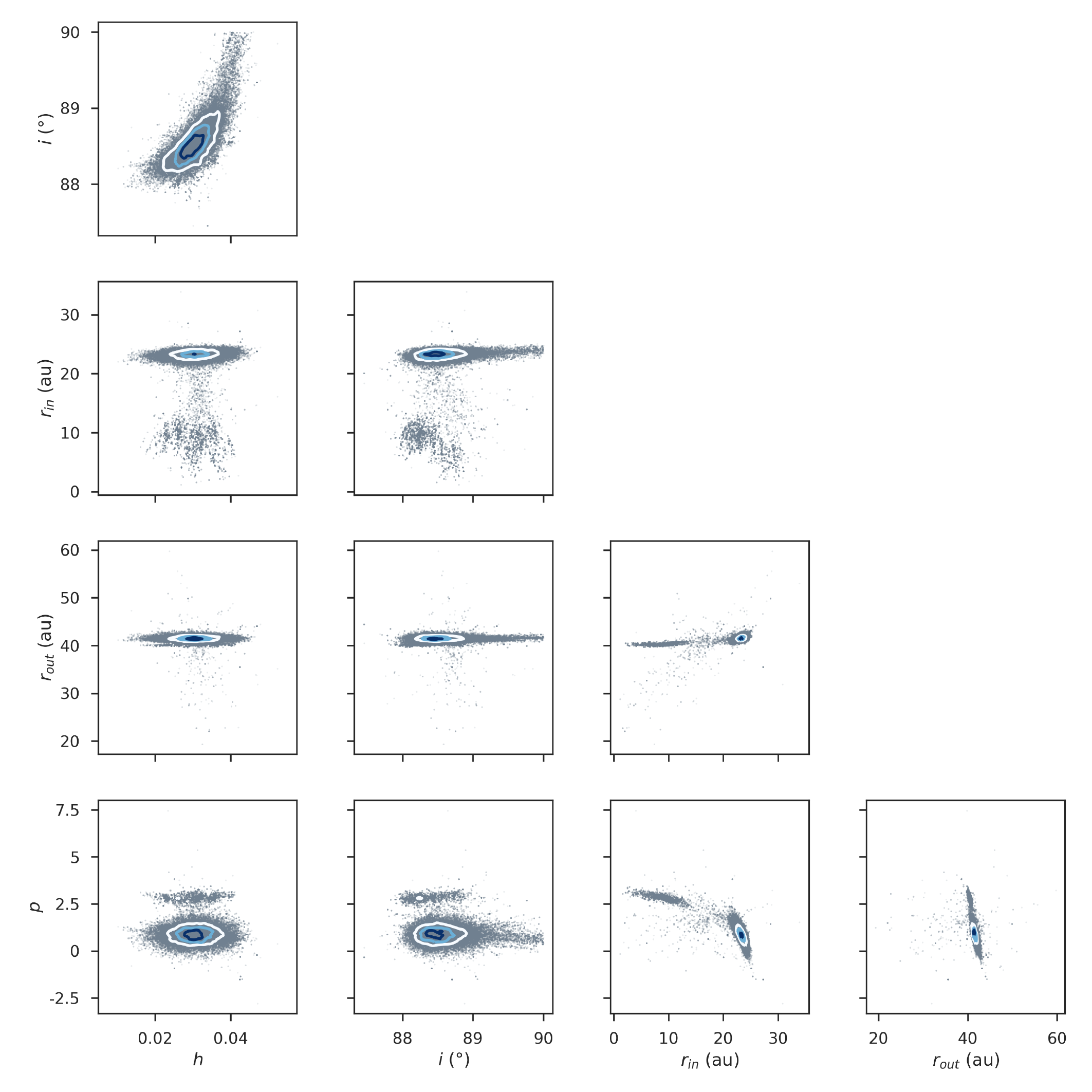}
  \caption{
  `Corner' plot showing degeneracies between a subset of the free parameters for the fiducial model.
  Two families of solutions are visible, and are most prominent in the $p$, $r_{in}$, and $r_{out}$ distributions.
  In addition, the two-dimensional slices through parameter space show limited degeneracies between radial structure, vertical structure, and viewing geometry.
  Of particular interest is the scale height $h$, which exhibits a noticeable degeneracy with the inclination $i$.
  }
  \label{fig: degeneracies}
\end{figure*}

\begin{figure*}[p]
  \centering

  \subfloat[Fiducial run best-fit model and residuals.]{%
    \includegraphics[width=0.87\linewidth]{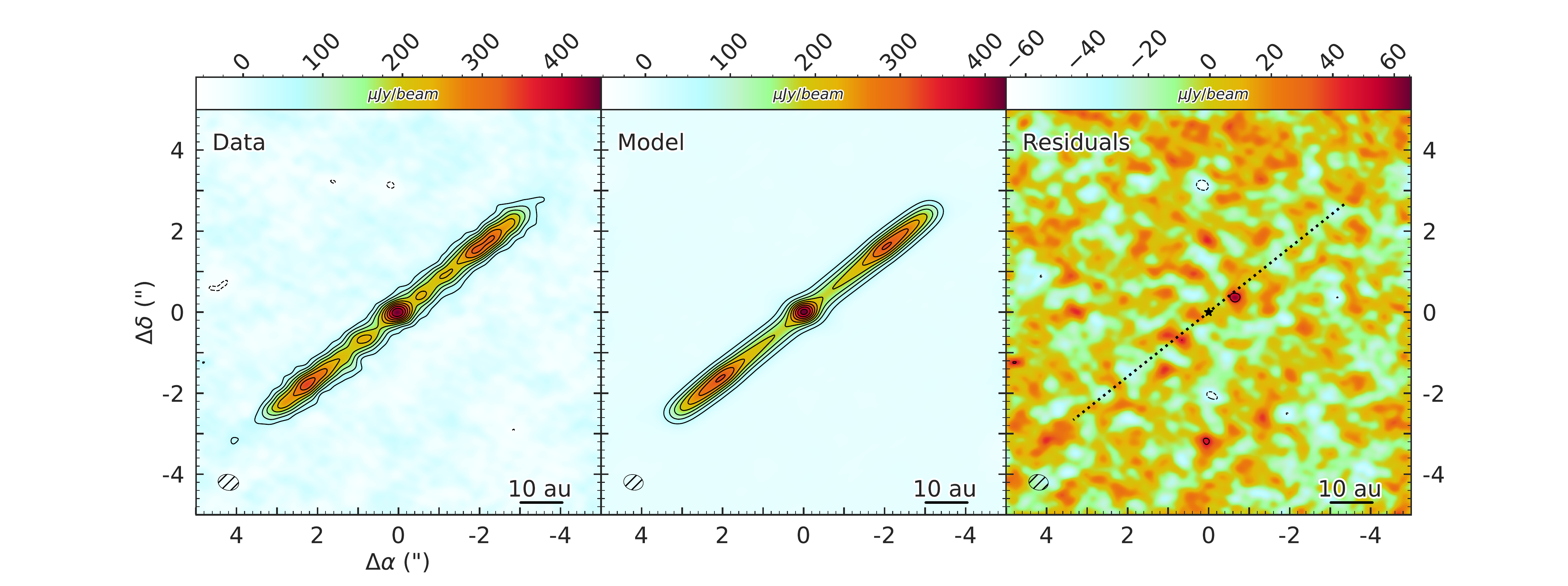}%
    \label{fig: fiducial}%
    }\qquad

  \subfloat[Ring run best-fit model and residuals.]{%
    \includegraphics[width=0.87\linewidth]{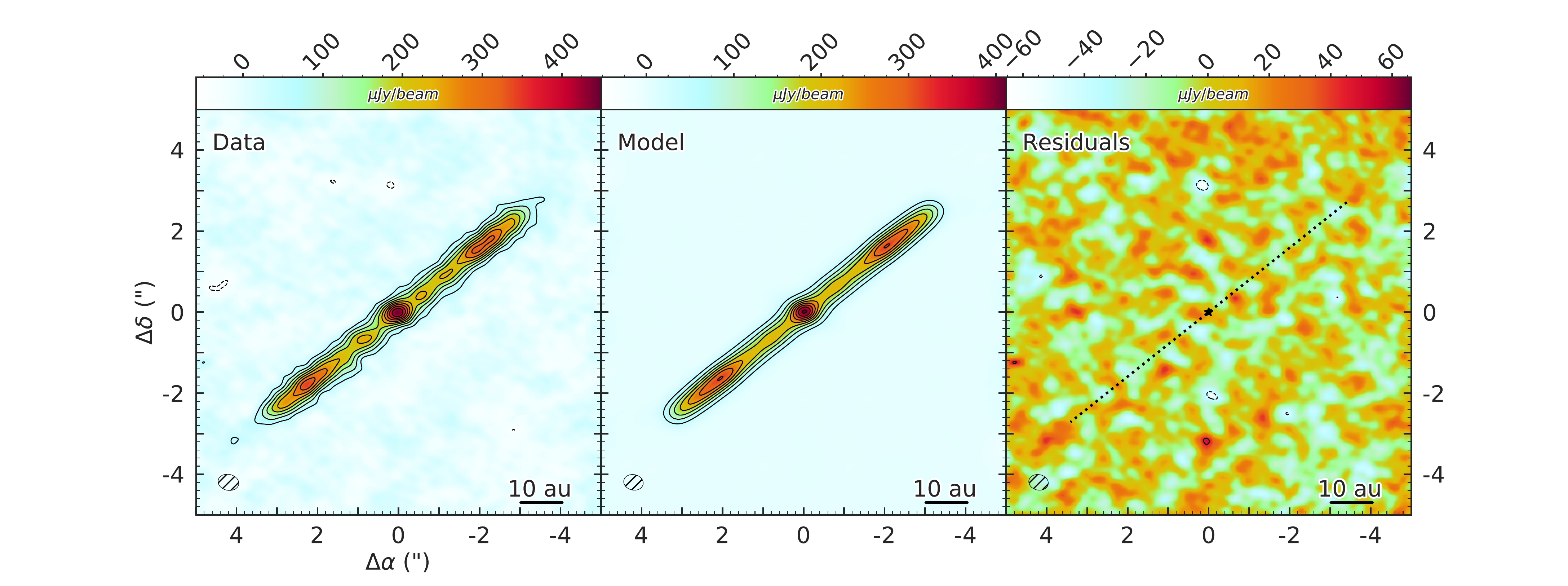}%
    \label{fig: ring}%
    }\qquad

  \subfloat[`Skinny' disk run best-fit model and residuals.]{%
    \includegraphics[width=0.87\linewidth]{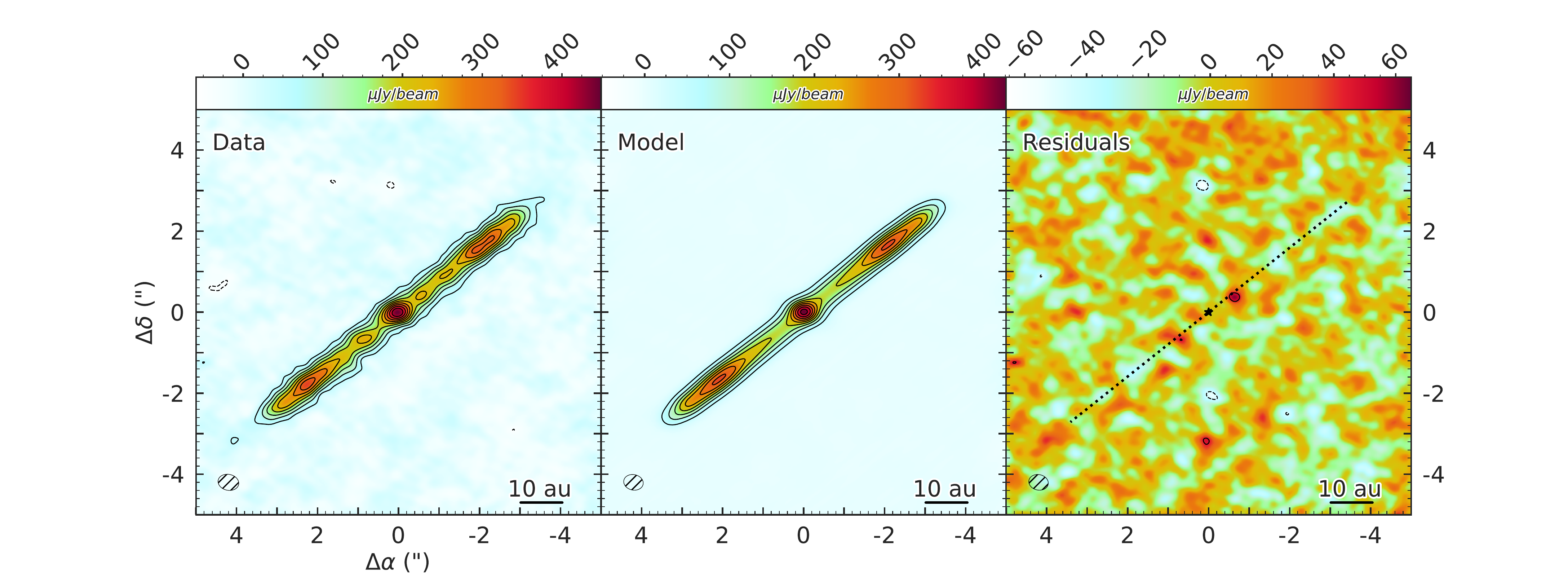}%
    \label{fig: skinny}%
    }\qquad

  \caption{Best-fit model image and residuals, sampled at the spatial frequencies of the ALMA data and cleaned using natural weighting. 
  Contours are integer multiples of the ALMA data $3\sigma$ confidence level. 
  In the residual maps the stellar location is marked with a star, and the disk extent and position angle are given by the dotted black line.
  For the fiducial model (a), $3\sigma$ residuals can clearly be seen at the at semi-symmetric positions $\sim \SI{10}{au}$ from the star. 
  Adding an additional ring of dust to the model (b) reduces these residuals by $\sim 25 \%$, but statistical analysis indicates that best-fit ring model does not conclusively describe the data better than the fiducial model.
  Fixing the scale height well below ALMA's resolution limits (c) results in a less statistically significant model, especially in the outer regions of the disk where ALMA is sensitive to the vertical structure of the disk.}
\end{figure*}

Marginalized posterior probability distributions for the fiducial model parameters are shown in Figure \ref{fig: kde}.
While the majority of the distributions are Gaussian, the three parameters that determine the radial structure of the disk ($p$, $r_{in}$, and $r_{out}$) exhibit slight bimodality. 
All three parameters are degenerate as can be seen from the `corner' plot in Figure \ref{fig: degeneracies}; in fact, the bimodality of the three parameters is a result of the existence of two distinct families of solutions in parameter space.
While the fiducial best-fit model has $p=0.9$, $r_{in}=\SI{23.3}{au}$, and $r_{out}=\SI{41.5}{au}$, a lower-likelihood family of solutions can clearly be seen in Figure \ref{fig: degeneracies}. 
The highest-likelihood model associated with this family has $p = 2.8$, $r_{in} = \SI{8.8}{au}$, and $r_{out} = \SI{40.3}{au}$.
More generally, mild degeneracies between the parameters determining vertical structure, radial  structure, and viewing geometry are visible, but the aspect ratio $h$ is correlated to a significant degree only with the inclination $i$.

The fiducial best-fit model image and residuals found in Figure \ref{fig: fiducial} provide further information as to the cause of the bimodal posterior distribution.
As can be seen from the residual map, the outer regions of the disk are reproduced well by the model; however semi-symmetric $3\sigma$ residuals, one `above' the disk midplane and one `below,' remain at projected stellocentric separations of $\sim \SI{10}{au}$. 
We note these residuals share the locations of the local intensity maxima described in \S \ref{section: results}. 
The convergence of these features leads us to consider the possibility of either a dust density enhancement (a ring) or reduction (a gap) in the inner regions of the disk. 
As a gap/ring would cause the radial surface brightness profile of the disk to deviate from that given by the simple power law used in our modeling, it could explain the bimodality in the posterior distributions of the parameters governing disk radial structure.

We first explored the effects of adding a gap  to the inner regions of the disk. 
The gap inner and outer radii were left as free parameters and the dust density within the gap was set to zero.
The gap consistently converged to regions where the dust density was already zero (interior to the disk inner radius or exterior to the disk outer radius); we take this as evidence that the data are not well characterized by a gap.
As a next step, we experimented with adding a ring to the disk.
The ring inner radius $r_{ring}$ was once again left as a free parameter and the ring width was fixed to \SI{0.1}{au}.
The ring was also characterized by a dust mass $M_{ring}$, which was evenly distributed across the radial extent of the ring. 
As can be seen in Figure \ref{fig: ring} this parameterization was better able to reproduce the `bump' in the radial surface brightness profile at $r \sim \SI{10}{au}$, reducing the best-fit model residuals by $\sim 25 \%$. 
However, $2 \sigma$ residuals are still present at the location of the bumps; the ring model's failure to accurately reproduce these features could be explained by the fact that the two local intensity maxima are not perfectly symmetric about the star, and together exhibit a position angle that deviates from that of the main disk. 
Image domain analysis indicates that the bump to the SE has a separation of $\sim \SI{10.2}{au}$, while the NW bump has a separation of $\sim \SI{7.6}{au}$ (Figure \ref{fig: boccaletti}, second pane). 
This discrepancy could be explained if the hypothetical ring were eccentric or if the two bumps were instead localized, asymmetric density enhancements within the outer ring.

The median values and best-fit model parameters for the fiducial and ring parameterizations can be found in Table \ref{tab: params}. 
We use both the AICc, a form of the Aikake Information Criterion (AIC) corrected for finite datasets, and the Bayesian Information Criterion (BIC) to compare goodness of fit between models with different numbers of free parameters.  
Both criteria attempt to strike a balance between overfitting and underfitting by rewarding models with higher probability and penalizing models with more free parameters; the BIC penalizes free parameters more severely than the AIC.
If the BIC scores of two models differ by $\Delta \text{BIC} < 2$ then neither model is significantly better or worse than the other, while $2 < \Delta \text{BIC} < 6$ implies `positive' evidence of a statistically improved fit.
If $6 < \Delta \text{BIC} < 10$ then `strong evidence' is implied, and $ \Delta \text{BIC} > 10$ implies `very strong' evidence.
The best-fit ring model is preferred to the fiducial model with $3.7 \sigma$ confidence on the AICc; conversely, the fiducial model is preferred to the ring model on the BIC with $\Delta \text{BIC} = 4.3$.
As such, we are not able to conclusively confirm the presence of an additional ring of mm dust grains in the AU Mic disk.

A model with a single $F_\star$ across all three dates was also investigated, but did not reproduce the data to the same degree of accuracy as the fiducial model.
Significant residuals were visible at the location of the star, and  a stellar flux density varying by more than a factor of 2 over a period of months to years is preferred with $7.3 \sigma$ confidence by the AICc and with $\Delta \text{BIC} = 35.0$.

\subsection{Investigating Vertical Structure}
\label{subsection: vertical analysis}

The posterior distribution for the aspect ratio $h$ suggests that the data are capable of measuring AU Mic's scale height despite the mild degeneracy between the aspect ratio and other parameters like the radial structure and viewing geometry.  
Even when marginalized over these other parameters, the posterior distribution indicates a measured value of $h=0.031 _{-0.004}^{+0.005}$, which translates to a $\sim 8 \sigma$ measurement of the scale height rather than an upper limit.
At the $\sim \SI{40}{au}$ outer edge of the disk, this aspect ratio implies a vertical scale height of \SI[parse-numbers=false]{1.24^{+0.20}_{-0.16}}{au}. 
The corresponding disk FWHM is \SI[parse-numbers=false]{2.9^{+0.5}_{-0.4}}{au}, which is consistent with the mean image-domain FWHM of $\sim \SI{2.8}{au}$ for projected separations $\leq \SI{40}{au}$ (Figure \ref{fig: boccaletti}; due to the edge-on inclination of the disk, the outer edge of the disk sets the apparent disk thickness at all projected separations).

To verify that we in fact measured the scale height, we investigate a model parameterization in which the scale height is set to a value well below ALMA's resolution limits.
The aspect ratio is fixed at a value of $0.003$, so that even at the outer edge of the disk the scale height is $\sim \SI{3}{\percent}$ of the beam size along the disk's vertical axis.
If the disk is in fact resolved by the observations, such a `skinny' disk model should perform significantly worse than the fiducial model.
Hence, we can quantify the statistical significance of our detection of the scale height by comparing the best-fit skinny model to the best-fit fiducial model.
The best-fit model image and residuals are shown in Figure \ref{fig: skinny}.
The skinny model results in a significantly poorer fit to the data than the fiducial model with variable aspect ratio, with the best fits differing at the $3.7 \sigma$ level according to the AICc and by a value of 5.8 on the BIC.

\section{Discussion}
\label{section: discussion}

Parametric modeling suggests that AU Mic's debris disk is nearly edge on, exhibits an increasing surface density with radius until $\sim\SI{42}{au}$, and reaches a maximum scale height of $\sim\SI{1.2}{au}$.
There is also marginal evidence for an additional annulus of dust at $r \sim \SI{10}{au}$.
Here we compare the results of our analysis with previous studies of AU Mic's debris disk.
Relevant quantities in the literature have been scaled by the new Gaia distance, although in most cases the scaling has no effect on the first significant figure.  
For cases where scaling by the Gaia distance changes any of the significant digits reported, the resulting value is referred to as ``Gaia-corrected.''

\subsection{Dust \& Gas Mass}
\label{subsection: dust mass}

MCMC fitting prefers a median dust mass of \\ \SI{9.28 \pm 0.13e-3}{M_\earth} for an opacity of \SI{2.3}{cm^2.g^{-1}}; this value tends to be slightly smaller than previous estimates, possibly because of the low disk flux reported in \S \ref{section: results} and because the new, shorter Gaia distance of \SI{9.725 \pm 0.005}{pc} should decrease the preferred dust mass.
We note that the reported uncertainties take into account neither the 10\% flux uncertainty nor assumptions about the dust opacity.
Assuming an opacity of \SI{2.7}{cm^2.g^{-1}} and a dust temperature of \SI{25}{K}, \citet{macgregor13} estimate a dust mass of $\sim \SI{0.01}{M_\earth}$ from a \SI{1.3}{mm} disk flux density of \SI[parse-numbers=false]{7.14^{+0.12}_{-0.25}}{mJy}.
\citet{matthews15} report an identical mass with 20\% uncertainty from a \SI{850}{\mu m} disk flux density of \SI{12.5 \pm 1.5}{mJy} and an opacity of \SI{1.7}{cm^2.g^{-1}}.
\citet{liu04} also take the opacity to be \SI{1.7}{cm^2.g^{-1}}, and report \SI{0.011}{M_\earth} from a disk flux density of \SI{14.4 \pm 1.8}{mJy} at \SI{850}{\mu m}.
Finally, \citet{strubbe&chiang06} calculate \SI{0.01}{M_\earth} by deriving a steady-state collisional cascade grain size distribution and fitting the disk's surface brightness and thermal spectrum to $V$- and $H$-band HST observations.

Detections of gas in debris disks are becoming increasingly common, and are changing our understanding of dust morphology in gas-bearing systems (see \S 4 of \citealp{hughes18}, and references therein).
Theoretical works such as that of \citet{takeuchi&artymowcz01} predict that small dust grain dynamics will be affected by the presence of gas for gas-to-dust ratios $\gtrsim 1$; Section 5.2 of \citet{hughes17} provides a discussion of the ways in which gas can interact with dust in debris disks.
As discussed in \S \ref{section: results}, even a large gas excitation temperature of \SI{250}{K} implies a $3 \sigma$ upper limit on the $^{12}$CO mass of only \SI{8.7e-7}{M_\earth}. 
Combined with our preferred dust mass of \SI{9.08e-3}{M_\earth} and assuming a conservative CO/H$_2$ ratio of $10^{-4}$, this CO mass corresponds to an upper limit of $\sim 1$ on the gas-to-dust ratio.
Such a low gas-to-dust ratio excludes the possibility of gas influencing the millimeter grain kinematics, and thus our measurement of the scale height.

\subsection{Disk Geometry}
\label{subsection: disk geometry}

The geometric properties of the disk inferred from our modeling generally agree with the literature. 
The median of the inclination posterior distribution $\ang{88.5}^{+0.3}_{-0.2}$ is marginally consistent with \cite{metchev05} who estimate $i \gtrsim \ang{89}$, but is not consistent with \cite{krist05} who report an inclination between \ang{89.43} and \ang{89.44}.
The median PA of $\ang{128.49} \pm 0.07$ falls within uncertainties of measurements by \citet{macgregor13} and \citet{krist05}; however, the litb-averaged PA interior to \SI{50}{au} from \citet{metchev05} is $\ang{129.8} \pm 0.2$, and \citet{schneider14} report an optical-wavelength PA of $\ang{127.8} \pm 0.2$ between \SIrange{50}{100}{au}.

\begin{figure}[t]
  \centering
  \includegraphics[width=\linewidth]{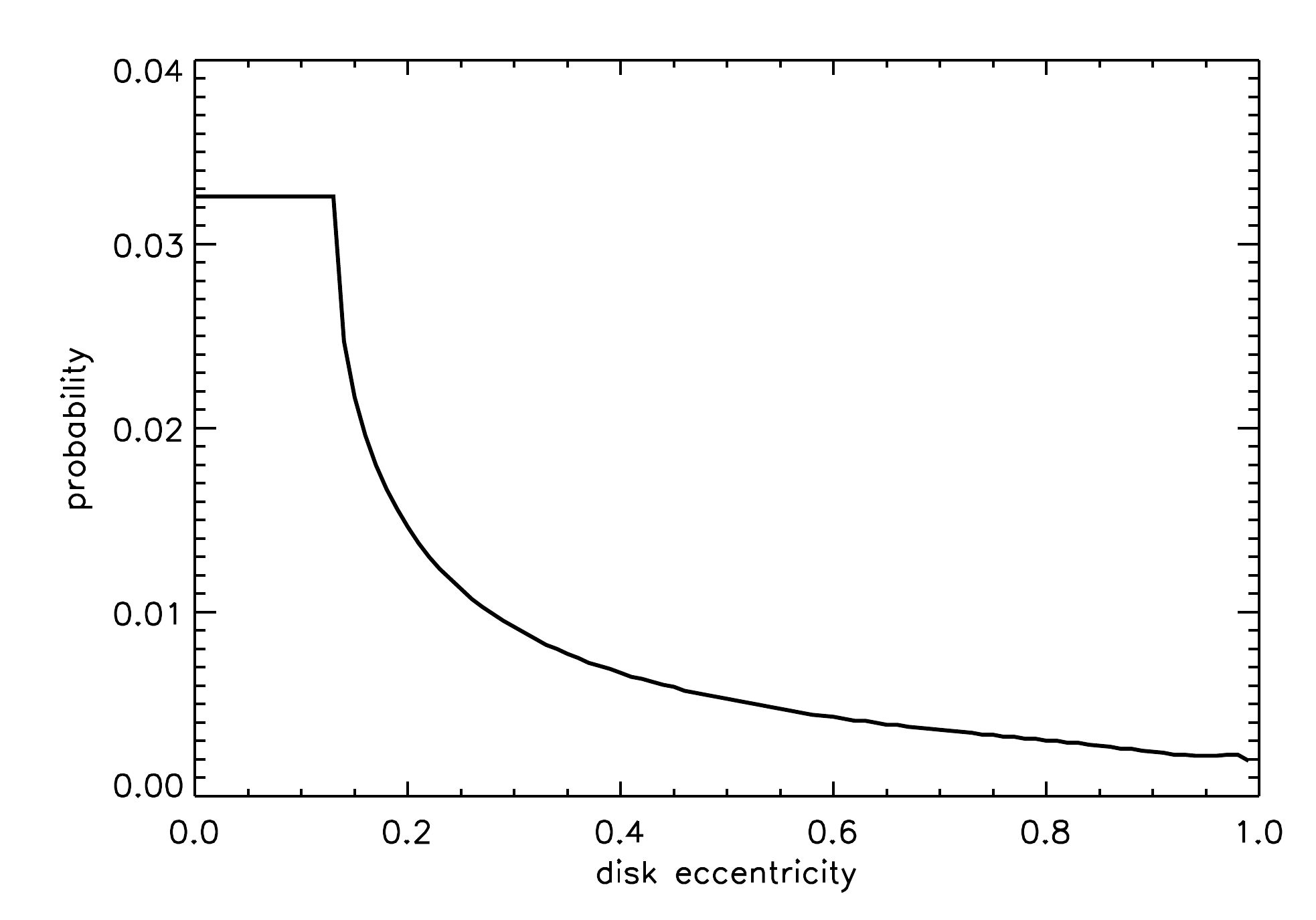}
  \caption{
    A plot of the probability density function (PDF) for the eccentricity of the AU Mic disk.
    The plot shows the probability density associated with a given eccentricity.
    The probability is based on the fraction of periapsis angles for a given eccentricity that are consistent with the observed SE-NW ansa flux density ratio: at low eccentricity any periapsis angle would produce the observed flux asymmetry, while at high eccentricity the line of apsides would need to be nearly aligned with the line of sight to produce the observed asymmetry.
    There is a 50\% chance that the eccentricity of is less than 0.15, a 65\% chance that it is less than 0.28, and a 95\% chance that it is less than 0.78.
  }
  \label{fig: eccentricity}
\end{figure}

The eccentricity of the AU Mic debris disk can be constrained from the SE-NW ansa flux density ratio $f_{\text{SE-NW}} = \frac{\SI{340 \pm 15}{mJy.beam^{-1}}}{\SI{330 \pm 15}{mJy \ beam}} = \num{1.030 \pm 0.065}$ reported in \S \ref{section: results}.
For a perfectly circular disk, the flux ratio for all pairs of points should be 1. 
For a disk with a nonzero eccentricity, the flux ratio will be largest between periapse and apoapse. 
When the disk's line of apsides (connecting periapse and apoapse) falls on the sky plane (perpendicular to the line of sight), $f_{\text{SE-NW}}$ is simply the apoapse-periapse ratio. 
If the line of apsides of the disk has any other orientation, $f_{\text{SE-NW}}$ is closer to unity than the apoapse-periapse ratio, and $f_{\text{SE-NW}}$ should equal one when the line of apsides points exactly along our line of sight.

For a disk with a small eccentricity, the fluxes at apoapse and periapse are still fairly close to one another and so the ratios for all orientations of the line of apsides fall within the measured $1\sigma$ confidence interval. 
The disk may also exhibit a larger eccentricity, such that the apoapse-periapse flux ratio is greater than the observed upper limit of $f_{\text{SE-NW}}=1.095$; then the disk must have its line of apsides oriented at an angle to the sky plane in order to satisfy the constraint on $f_{\text{SE-NW}}$. 
As the disk eccentricity increases further, the line of apsides must lie closer and closer to our line of sight for $f_{\text{SE-NW}}$ to fall within the observed range.
Assuming all edge-on disk orientations are equally likely, higher eccentricities are thus less likely given the observed fluxes. 
On the other hand, all eccentricities below 0.13---the value at which the apoapse-periapse flux ratio is equal to the observed upper limit of $f_{\text{SE-NW}}=1.095$---are equally likely. 

Figure \ref{fig: eccentricity} shows the probability distribution of the disk's eccentricity given the observed value of $f_{\text{SE-NW}}$.
We assume a disk radius of \SI{40}{au}, and a stellar mass, effective temperature, and luminosity of \SI{0.5}{M_\sun}, \SI{3500}{K} \citep{plavchan09}, and \SI{0.09}{L_\sun} respectively. 
The grain size distribution is determined by a power law index $q=3.5$, and grain absorptivity scales as $\lambda^{-1}$ for wavelengths larger than the grain size.
We find that there is a 50\% probability that the eccentricity of the AU Mic debris disk is less than 0.15, a 65\% probability that it is less than 0.28, and a 95\% probability that it is less than 0.78.
Section 5.4 and Figure 7 of \citet{wyatt99} contains a similar discussion of the degeneracy between periapse orientation and eccentricity in relation to apoapse-periapse flux ratios.

\subsection{Radial Structure}
\label{subsection: radial discussion}

\begin{figure}[t]
  \includegraphics[width=\linewidth]{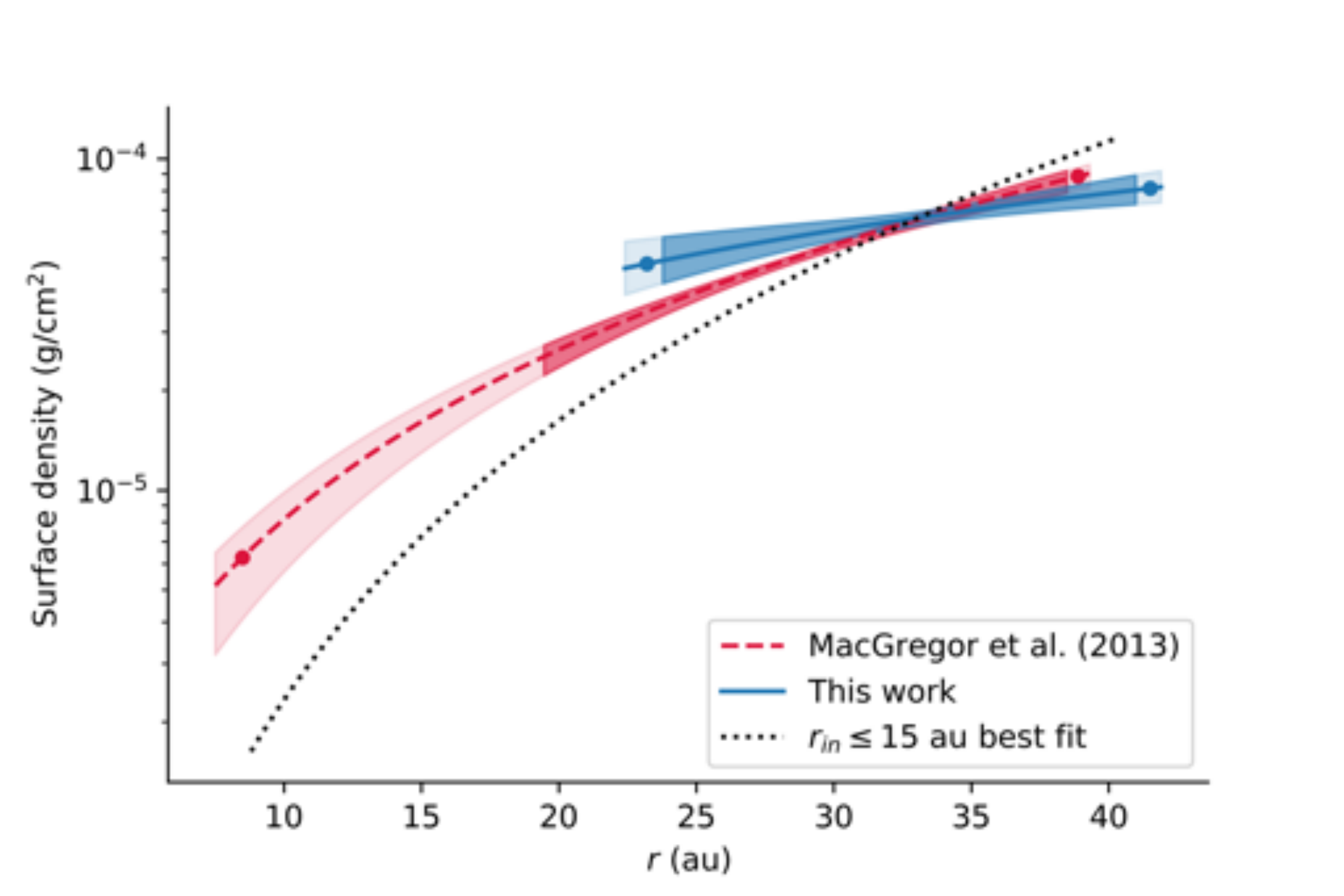}
  \caption{
    Comparison between the best-fit surface density profile obtained in \citet{macgregor13} (dashed red line) and the median surface density profile from this work (solid blue line).
    Also included is the best-fit surface density profile associated with the lower-likelihood family of solutions (dotted black line).
    The vertical extent of the shaded regions marks $1\sigma$ confidence intervals on the surface density. 
    The points designate the best-fit inner and outer radii for each model, and the horizontal extent of the more transparent shaded regions show $1 \sigma$ confidence intervals on the inner and outer radii.
    Not included in the uncertainties are the 10\% flux uncertainties of each observation.
    }
  \label{fig surface_density}
\end{figure}

Modeling of the radial structure discussed in \S \ref{subsection: radial analysis} yields an inner radius $r_{in} = \SI[parse-numbers=false]{23.2^{+0.6}_{-0.8}}{au}$, an outer radius $r_{out} = \SI[parse-numbers=false]{41.5^{+0.5}_{-0.6}}{au}$, and a power law exponent $p=0.9_{-0.4}^{+0.5}$.
As can be seen in Figure \ref{fig surface_density}, these results are broadly consistent with the previous analysis of millimeter wavelength emission from the disk performed by \citet{macgregor13}, especially in the high-SNR region of the disk.
However, in addition to the preferred solution quoted above, the MCMC posterior distributions are suggestive of a second family of solutions at $r_{in} \sim \SI{9}{au},\ r_{out} \sim \SI{40}{au},\ p \sim 2.8$ (Figures \ref{fig: kde} \& \ref{fig: degeneracies}).
This lower-likelihood solution is generally consistent with the Gaia-corrected best-fit radial structure quoted by \citet{macgregor13}: $r_{in} = \SI[parse-numbers=false]{8.5_{-1.0}^{+11.0}}{au},\ r_{out} = \SI{38.9 \pm 0.4}{au},\ p = 2.3_{-0.3}^{+0.2}$ (Figure \ref{fig surface_density}).
The inner radius is poorly constrained in both \citet{macgregor13} and this work: the former report a Gaia-corrected $3 \sigma$ upper limit of $\sim$\SI{20}{au}, whereas conversely we find a $3 \sigma$ lower limit of $\sim \SI{5}{au}$.

The lower-likelihood solution is probably associated with the local intensity maxima located at stellocentric separations of $\sim \SI{7.6}{au}$ to the NW and $\sim \SI{10.2}{au}$ to the SE.
Our exploratory modeling of these features in \S \ref{section: analysis} raises the possibility of an annulus of dust interior to the main disk---if such an annulus does exist, the $\sim 2$ times lower resolution observations from \citet{macgregor13} would probably not have been able to distinguish the annulus emission from that of the main disk.
In fact, an unresolved annulus would likely bias the inner radius preferred by the authors' modeling towards smaller values and could account for the differences in our characterization of the radial structure.

While the annulus model provides only a marginally significantly improved fit to the data, previous observations spanning a wide range of wavelengths have recovered surface brightness enhancements at the same projected stellocentric separation ($\sim \SI{10}{au}$) as the hypothetical annulus.
A local maximum is present to the SE of the star at this separation in lower-resolution ALMA observations by \citet{macgregor13}, and there is a suggestive peak in the noise on the opposite side of the star as well.
Although these surface brightness enhancements do not attain $3 \sigma$ significance after subtraction of an axisymmetric model, the independent detection of these features in both data sets suggests that they may be real.  
\citet{schneider14} also observe an optical-wavelength `bump' $\sim \SI{13}{au}$ SE of the star and slightly elevated from the disk midplane (i.e. to the NE).
No matching surface brightness enhancement is observed on the NW side of the disk, which is obscured by the STIS occulting wedge for $r \lesssim \SI{12}{au}$.
On the other hand, the scattered light emission on the NW side is not symmetric about the midplane and the authors tentatively identify a warp below the midplane extending to a Gaia-corrected distance of $\sim \SI{43}{au}$ from the star.
They note that these features may share a common cause, and posit that dust orbits in the inner disk are non-coplanar with those found at larger separations.

Near-infrared GPI observations presented in \citet{wang15} further corroborate the presence of a `bump' to the SE, characterized by a FWHM roughly triple the FWHM at an equivalent separation on the NW side of the disk. 
No features are detected at a corresponding separation to the NW.
Figure 4 of \citet{wang15} shows a composite map of the bump as seen by GPI, STIS, and ALMA; the common location of the bump in both scattered light and mm observations would indicate that the mm- and micron-sized grains are cospatial.
The micron-sized grains traced by scattered light, and all larger bodies ranging up to the cm-to-m sizes characterizing the top of the collisional cascade, are thought to originate from a narrow `birth ring' at $\sim \SI{40}{au}$ \citep{strubbe&chiang06}. 
Thus it is possible that all features are located at a true stellocentric separation of $\sim \SI{40}{au}$, with the apparent stellocentric separation of $\sim \SI{10}{au}$ being due to high-inclination projection effects.
If the two features seen in our data were located at \SI{40}{au}, the NW feature would exhibit an angle of $\sim \ang{10}$ with respect to the line of sight, while the SE feature would exhibit an angle of $\sim \ang{14}$.
In principle, the true stellocentric separation could be constrained by the range of temperatures allowed by AU Mic's SED. 
However such an analysis would be complicated by the relatively small mass contribution of the feature, the need to incorporate constraints on the small grains from scattered light imaging, and assumptions about the grain size distribution.

Planets are often invoked to explain rings in debris disks; P.~Plavchan et al.~(in preparation) propose a Jovian-mass exoplanet candidate interior to \SI{1}{au}, but AU Mic's stellar activity makes it difficult to confirm the radial velocity detection.
It is unlikely that a planet so close to the star would be responsible for a ring at $\sim\SI{10}{au}$.
On the other hand, a planet might be related to AU Mic's fast-moving features in the point-source parent body picture of \citet{sezestre17}.
The $< \SI{1}{au}$ separation of the potential planet detection by Plavchan et al. is far interior to the preferred \SI{8}{au} separation that \citet{sezestre17} derive to explain the origin of the fast-moving features, but separations $< \SI{5}{au}$ are not ruled out by the latter's analysis.

\subsection{Vertical Structure}
\label{subsection: vertical discussion}

\afterpage{
  \begin{figure*}
    \centering
    \caption{}
    \label{fig: scale height comparison}
    \includegraphics[width=\linewidth]{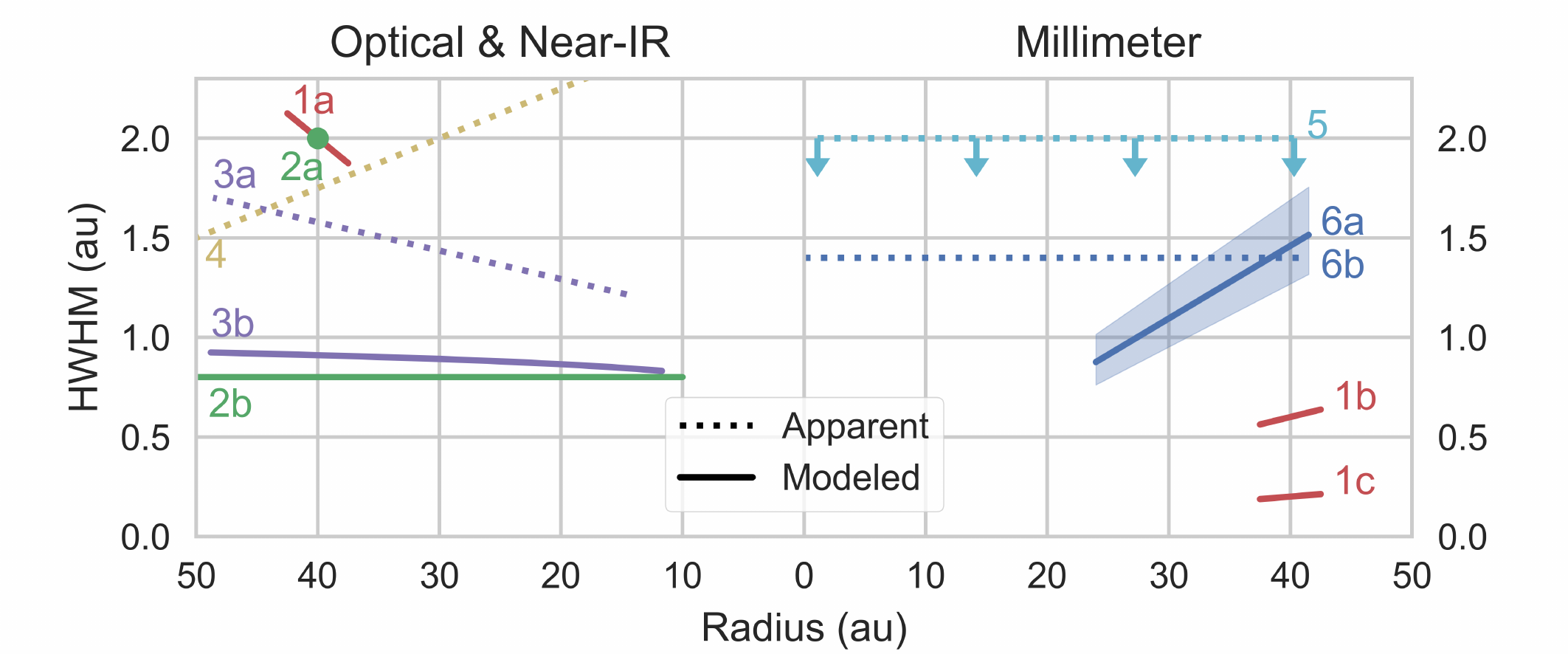}
  \end{figure*}
  \begin{table*}
    \centering
    \begin{tabular*}{\textwidth}{l @{\extracolsep{\fill}} cccc}
      \toprule
      Label & Reference & Wavelength & HWHM at \SI{40}{au} (au) & Methodology \\
      \midrule
      1a & \citet{schuppler15} & $ < \SI{70}{\mu m}$  & 2    & Collisional model \\
      1b & --- & \SI{0.445}{\mu m}--\SI{1.3}{mm}      & 0.6  & --- \\
      1c & --- & $> \SI{100}{\mu m}$                  & 0.2  & --- \\
      2a & \citet{metchev05}   & \SI{1.63}{\mu m}     & 2    & Apparent  \\
      2b & --- & \SI{.0647}{\mu m}--\SI{1.63}{\mu m} & 0.8  &  Model($i=\ang{89}$) \\
      3a & \citet{krist05}     & \SI{.0430}{\mu m}--\SI{0.0833}{\mu m} & 1.5 & Apparent \\
      3b &     ---             &         ---           & 0.9  & Model ($i=\ang{89.4}$) \\
      4  & \citet{schneider14} & \SI{.2}{\mu m}--\SI{1.05}{\mu m} & 1.2 & Apparent \\
      5  & \citet{macgregor13} & \SI{1.28}{mm} & $\leq 2$     & Apparent \\
      6a & This work           & \SI{1.35}{mm} & $1.46^{+0.24}_{-0.19}$ & Model ($i=\ang{88.5}$) \\
      6b &       ---           &   ---         & 1.4         & Apparent \\
      \bottomrule
    \end{tabular*}
    \caption{Measurements of the vertical structure of AU Mic's debris disk.}
    \label{tab: scale height comparison}
  \end{table*}
}

As discussed in \S \ref{subsection: vertical analysis}, our MCMC analysis yields a median aspect ratio of $h=0.031 _{-0.004}^{+0.005}$; at a reference radius of \SI{40}{au}, this translates to a vertical scale height $H = \SI[parse-numbers=false]{1.24^{+0.20}_{-0.16}}{au}$ and a FWHM of \SI[parse-numbers=false]{2.9^{+0.5}_{-0.4}}{au}.
The FWHM is thus only $\sim 2/3$ the size of the combined-data beam projected onto the vertical axis of the disk.
While it may seem improbable that we measure a vertical thickness below the image-domain spatial resolution of the combined data, our results can be explained by the following observations.
First, the beam size of the naturally weighted combined image ($\ang{;;0.52} \times \ang{;;0.39}$) is not wholly representative of the resolution of the data. 
The smallest naturally weighted beam FHWM from the three individual observations is $\ang{;;0.30} = \SI{2.9}{au}$ (Table \ref{tab:observations}), while the spatial scale traced by the longest baseline is $\ang{;;0.22} = \SI{2.1}{au}$.
Second, the scale height $H$ refers to the disk height as measured from the midplane; the observable quantity that must be resolved in order to measure the scale height is actually the total vertical thickness $2H$.
Third, the scale height represents the standard deviation of a Gaussian distribution of dust particles that in fact reaches well beyond the extent of the scale height.
For example, considering that the peak signal-to-noise (SNR) of the data is $\sim 23$ and that the vertical distribution of dust is assumed to be Gaussian, the SNR should remain above 3 over a total vertical extent of $\sim$\ang{;;0.35} (\SI{3.4}{au}). 
The combination of these factors indicates that it is plausible that we are able to detect a scale height smaller than the resolution of the image-domain data.

Our ability to measure the disk's vertical thickness could also be affected by uncalibrated phase noise, which could cause the true angular resolution of the data to be worse than that given by the beam.
\citet{boehler13} find that introducing Gaussian-distributed phase noise to synthetic ALMA observations of highly-inclined {protoplanetary} disks has little effect on the measured scale height value, although the corresponding uncertainties do grow.
The authors also find that introducing phase noise dramatically increases model reduced $\chi^2$ values; our low reduced $\chi^2$ value of {1.032} provides an indirect indication that phase noise does not affect our results.
To further investigate the effects of phase noise on the resolution, we imaged the two test quasars included in the ALMA data with the same clean parameters that were used to image the disk.
We find that the beam-convolved FWHMs of the quasars observed during the two long-baseline observations are slightly larger than the corresponding beam FWHMs; however, there are several reasons to believe that the phase noise does not prevent us from resolving the vertical structure of the AU Mic disk. 
The beam-deconvolved FWHMs of the test quasars projected onto the vertical axis of the disk ($\sim$ \ang{;;0.10} and \ang{;;0.16}) are still smaller than the mean beam-subtracted vertical disk FWHM (\ang{;;0.28}).  
Furthermore, since the quasar was observed only every other loop, the $u$-$v$ coverage and SNR are poorer than the full AU Mic data set, suggesting that the measured phase noise is an upper limit on that of the AU Mic data.  
We therefore continue to interpret our result as a resolved measurement of vertical structure, but if the phase noise is at the upper end of the range suggested by the quasar measurements, it may be an upper limit instead.

AU Mic's vertical structure has been previously measured in the optical and near-IR \citep[e.g.][]{krist05,metchev05}; this work's goal of measuring the disk's millimeter-wavelength vertical structure is partially motivated by the theoretical work of \cite{thebault09}, who predicts that the observed scale height of a debris disk should increase for smaller observing wavelengths.
\citet{thebault09} suggests that the smaller grains traced by short-wavelength observations can be placed on inclined orbits by radiation pressure (and to first order, disk winds) even in the absence of large bodies dynamically stirring the disk. 
This effect preferentially excites smaller dust grains, `puffing' up the disk at mid-IR to visible wavelengths, while the larger grains that dominate emission at longer wavelengths remain near the midplane.
\citet{thebault09} proposes a `natural' minimum debris disk thickness due to stellar winds and radiation pressure of $h = \num{0.04 \pm 0.02}$ as seen at wavelengths smaller than \SI{50}{\mu \meter}.
\citeauthor{thebault09} also runs a collisional model tailored to the AU Mic system assuming no intrinsic dynamical excitation, and reports $H = \SI{1}{au}$ for $r \leq \SI{40}{au}$ when degraded to the resolution of scattered light images.
Although this scale height falls within the range of `natural' debris disk thicknesses, the \citeauthor{thebault09} stresses that this does not amount to an assertion that the disk is dynamically cold due to the simplicity of their model and fitting process.

In light of the work of \citet{thebault09}, the wave\-length-dependence of AU Mic's scale height is of particular interest.
Our measured HWHM of \SI[parse-numbers=false]{1.46^{+0.24}_{-0.19}}{au} at $r=\SI{40}{au}$ is marginally consistent with values cited in the literature derived from observations spanning the optical to the sub-millimeter, ranging from roughly \SIrange{0.2}{2}{au} (Figure \ref{fig: scale height comparison}, Table \ref{tab: scale height comparison}).
\citet{schuppler15} place an upper limit of \SI{0.05}{\radian} on the \SI{1.3}{mm} semi-opening angle (equivalent to the aspect ratio for small angles) by extracting the image-domain vertical profile from vertically unresolved ALMA observations presented in \citet{macgregor13}, and estimate a visible-wavelength semi-opening angle of \SI{0.03}{\radian} by reading off the vertical scale height from STIS observations of the disk presented in \citet{schneider14}.
\citet{metchev05} approximate a $H$-band Keck FWHM of $\sim \SI{4}{au}$ at a separation of \SI{40}{au}; \citet{krist05} fit a vertical Lorentzian profile to multicolor HST observations of the disk and find the Gaia-corrected FWHM interior to \SI{50}{au} to fall between \SIlist{2.4;3.4}{au}.

Because the measurements quoted above are determined from the observed vertical thickness of AU Mic's disk, they can be affected by the radial structure and viewing geometry of the disk as well as scattering effects. 
Parametric modeling, which accounts for such effects, thus provides a more reliable way to assess AU Mic's vertical structure.
For example, \citet{krist05} report a FWHM between \SIlist{1.73;1.74}{au} at $r=\SI{20}{au}$ and an inclination of \ang{89.4} from three-dimensional scattering models.
\citet{metchev05} find that a constant scale height of \SI{0.8}{au} adequately reproduces the observed mean disk thickness assuming an inclination of \ang{89}.
Collisional modeling can also be used to learn about AU Mic's vertical structure: collisional velocities---and thus dust production---are affected by the maximum eccentricity $e_{max}$ of planetesimal orbits, which in turn can be related to the disk vertical stucture (see \S \ref{inferring mass} below). 
\citet{schuppler15} perform such modeling constrained by photometric observations spanning visible to millimeter wavelengths and quote a reference model opening angle of \SI{0.015}{rad}.
The authors note that an opening angle of \SI{0.005}{rad} better reproduces the disk spectral energy distribution (SED) in the long-wavelength regime beyond \SI{100}{\mu m}, but is unable to reproduce flux measurements for $\lambda \leq \SI{70}{\mu m}$. 

In sum, estimates of AU Mic's vertical structure vary depending on the wavelength of observation, the techniques used, and the assumptions made.
Rigorous comparison between measurements in the literature is difficult---for example, \citet{krist05} assume a flared vertical profile while other authors use a linear parameterization for the scale height.
Nevertheless, some general statements may be made.
HWHM values determined from optical and near-infrared observations range from roughly \SI{0.8}{au} to \SI{2}{au} at a radius of \SI{40}{au}, while the HWHMs determined at least in part from mm observations range from \SI{0.2}{au} to \SI{2}{au}, the latter being an upper limit. 
While our measurement of the scale height exceeds measurements by \cite{krist05} and \cite{metchev05} in the optical and near-IR (in apparent contradiction to the predictions of \cite{thebault09}), we note that both authors report higher inclinations than our $i=\ang{88.5}^{+0.3}_{-0.2}$.
Due to the degeneracy between scale height and inclination (e.g. Figure \ref{fig: degeneracies}), adopting a higher inclination would lead to a smaller estimate of the scale height.
We conclude that the two wavelength regimes do not provide radically different values, and we cannot unequivocally confirm or deny the prediction of \cite{thebault09}.
These observations provide an important datapoint for the vertical height of large particles in the cascade from which to consider how this connects to the distribution of smaller particles. 
If the vertical height is found to decrease with particle size (as might be inferred from some estimates of the short wavelength vertical height) this could point to an increased role for damping with decreasing particle size, perhaps due to the increased strength of such particles {\citep[e.g.,][]{housen&holsapple90}}.
This work provides a parent belt vertical height measurement needed for modeling of this effect.


In recent years, the vertical structure of several other debris disks has been tentatively resolved with ALMA.
Observations of CO~$\mathrm{J}=3-2$ and $2-1$ line emission from the edge-on $\beta$ Pic debris disk at a spatial resolution of $\sim \SI{5.5}{au}$ presented by \citet{matra17} suggest that the disk is resolved in the vertical direction.
The beam-subtracted apparent FWHM, determined in a similar manner as the bottom pane of Figure \ref{fig: boccaletti} in this work, ranges between $\sim$\SIrange[range-phrase=\ and\ ]{7}{12}{au} over the $\sim \SI{100}{au}$ radial extent of the disk.
Assuming Keplerian rotation and an edge-on inclination, the authors report the following scale height-radius relation:
\begin{align}
    H = 7.0 \pm 0.6 \times \qty(\frac{R}{\SI{85}{au}})^{0.75 \pm 0.02} \si{au}
\end{align}
where the scale height $H$ is the standard deviation of the Gaussian vertical density distribution.
For reference to our work, this corresponds to a scale height of \SI{4}{au} at a separation of \SI{40}{au}.

As pointed out in \S 4.9 of \citet{marino16}, the scale height of a narrow ring may be resolved even if the ring is not close to edge-on.
The authors estimate that for $\sim \ang{;;0.25}$ resolution observations of the low-inclination ($i \approx \ang{30}$) HD 181327 debris disk at a SNR of $\sim 50$, an aspect ratio of $h=0.083$ could be constrained to within $\pm 0.005$. 
Along the same lines, \citet{kennedy18} analyze $\SI{11.6}{au} \times \SI{13.1}{au}$ spatial resolution observations of the moderately-inclined ($i \approx \ang{77}$) HR 4796A debris disk, reporting a marginal detection of the disk scale height.
The authors find that a vertically resolved ring (FWHM $= \SI{7 \pm 1}{au}$) is favored over a vertically unresolved ring (FWHM $< \SI{4}{au}$) with $\Delta BIC = 6.8$, meeting the condition for `strong' evidence of a statistically higher-quality fit ($\Delta BIC > 6$). 
This FWHM corresponds to a Gaussian standard deviation $H \approx \SI{3.0 \pm 0.4}{au}$ and, at the $\sim \SI{80}{au}$ radial location of the ring, a scale factor of $h \approx 0.038 \pm 0.005$.
In sum, the vertical structure of debris disks is just beginning to be resolved in the millimeter thanks to the improvements in sensitivity and resolution provided by ALMA; to our knowledge, AU Mic exhibits the narrowest millimeter-wavelength scale height of any known debris disk.


\subsection{Inferring the Mass of Stirring Bodies}
\label{inferring mass}

Information regarding the bodies dynamically stirring AU Mic's disk can be recovered by relating the scale height to the dynamical excitation of the disk's dust grains.
As discussed by \citet{thebault09} and \citet{quillen07}, the planetesimals responsible for stirring the disk impart kinetic energy to the dust, perturbing them from a Keplerian orbit and thus increasing their orbital eccentricity dispersion $\langle e^2 \rangle$. 
Here we define $\bar{e} = \sqrt{\langle e^2 \rangle}$ and $\bar{i} = \sqrt{\langle i^2 \rangle}$, where $\langle i^2 \rangle$ is the inclination dispersion of the dust grain orbits.
In equilibrium there is an equipartition between the vertical and in-plane components of the velocities imparted to the grains, so $\bar{i} = {\bar{e}}/{2}$.
Inclination can be related to the observed FWHM using $\bar{i} = \sqrt{2}$FWHM for small angles; thus inclination is related to our Gaussian-standard-deviation aspect ratio by $\bar{i} \approx 2.355 h / \sqrt{2}$.
The interparticle relative velocity $\langle v_{rel} \rangle$ can then be determined directly from observables using the following relation \citep{wetherill&stewart93,wyatt&dent02}:
\begin{gather}
  \langle v_{rel} \rangle \approx v_{Kep}(r) \sqrt{\bar{i}^2 + 1.25 \bar{e}^2} \approx 2.355 \sqrt{3} v_{Kep}(r) h
\end{gather}
where $v_{Kep}(r)$ is the Keplerian velocity at radius $r$. 
Adopting a stellar mass of \SI{0.5}{M_\sun} and taking $r = \SI{40}{au}$, $h=0.031_{-0.004}^{+0.005}$ yields $v_{rel} = \SI[parse-numbers=false]{420^{+70}_{-50}}{m/s}$.

The velocity dispersion of the dust grains will be excited to about the escape velocity of the largest bodies governing the disk dynamics in the absence of any significant damping of their velocity dispersion. 
This result arises because viscous stirring has a larger cross section (i.e. shorter characteristic timescale) than collisions as long as the velocity dispersion is less than the escape velocity of the largest bodies that dominate the stirring. 
The two cross sections (and timescales) become comparable as $v_{rel}$ approaches $v_{esc}$ and collisions start to dominate, limiting the growth of $v_{rel}$ to about $v_{esc}$ \citep[e.g.,][]{schlichting14}.
As such, we can use our estimate for $v_{rel}$ to place a lower limit on the escape velocity $v_{esc}$ and thus size $a_{big}$ of bodies stirring the disk.
Assuming a typical asteroid density of \SI{2}{\g.\cm^{-3}} \citep{carry12}, we find $a_{big} \sim \SI[parse-numbers=false]{400^{+60}_{-50}}{km}$ and $m_{big} \sim \SI[parse-numbers=false]{5.3^{+2.6}_{-2.0} \times 10^{20}}{kg}$, i.e. $\sim 4\%$ the mass of Pluto.

On the other hand, the disk may be in a steady state in which velocity damping is balanced with excitation (rather than damping being inefficient).
Under this condition, the scale height provides a joint constraint on the number and mass of stirring bodies rather than their size.
To infer such this constraints from the observed disk scale height, we refer to theoretical models of steady state size-dependent velocity distributions in the collisional cascade desribed in \citet{pan&schlichting12}.
We assume a disk in steady state in which velocity excitation (``stirring'') occurs via close encounters with large planetesimals and velocity damping occurs via direct collisions with other (small) bodies. 
Because the disk is assumed to be in steady state, we can equate the expressions for stirring and damping rates and solve for the velocity of dust grains. 
The computation equating the stirring and damping is analogous to that of Equations 15 \& 16 in \citet{pan&schlichting12}; however, while Equations 15 \& 16 assume that catastrophic collisional destruction dominates the velocity damping, we assume that \textit{all} collisions with smaller bodies contribute to damping.
We assume an eccentricity of 0.03 {(typical of the major planets of the Solar System)} for the perturbing bodies in order to determine the velocity dispersion $v(R)$ of the largest bodies and relate the dust grain velocity to the disk scale height as described above.
Assuming a HWHM of \SI{1.5}{au} at \SI{40}{au}, a stellar mass of \SI{0.5}{M_\sun}, and a dust mass of \SI{9.28e-3}{M_\earth}, these models give the following joint constraint on the number and mass of perturbing bodies: 
\begin{equation}
    \sqrt{N_p} m_{p} \sim \SI{1.8}{M_\earth} 
\end{equation}
The maximum perturbing body mass occurs when $N_p = 1$; a single planet would then have a mass of $\sim \SI{1.8}{M_\earth}$ and a radius of $\sim \SI{1.2}{R_\earth}$, assuming a mean density of \SI{5.5}{\g.\cm^{-3}} characteristic of Earth. 
{For eccentricities slightly higher than the chosen $e=0.03$, this constraint should scale linearly with eccentricity.}

In sum, the largest bodies in the AU Mic disk cannot be smaller than \SI[parse-numbers=false]{400^{+60}_{-50}}{km} or they would not be able to stir the dust grains to the velocity dispersion inferred from the measured scale height.
Conversely, the most massive body in the disk cannot be larger than $\sim \SI{1.8}{M_\earth}$ or the scale height would exceed the value measured in this work.
Our conclusions regarding the stirring bodies are probably limited to the outer region of the disk, especially in the case of an ensemble of \SI{400}{km} planetesimals.
If scale height increases with radius, the observations are likely only sensitive to the scale height near the $\sim \SI{40}{au}$ outer disk radius (where the surface density and thus SNR is highest) because the maximum scale height is comparable to the spatial resolution of the data.
Bodies external to the disk are also capable of stirring the disk at a distance, through secular and/or mean-motion resonant perturbations; consideration of external perturbers is deferred to future work.
If external stirring is significant, then the above constraints on the properties of embedded (non-external) perturbers are all upper limits.

The properties of the perturbers estimated in this work are consistent with those inferred independently by \citet{chiang&fung17}.  
In their scenario, the catastrophic disruption of a Varuna-sized progenitor and the ongoing clearing of debris from this event underlie the fast-moving infrared features observed by \citet{boccaletti15,boccaletti18}.\footnote{
    {As \citet{chiang&fung17} discuss, the avalanche scenario is not without its problems, among them the ad hoc and unproven assumption that the star AU Mic emits a wind that varies on a ten-year timescale. 
    In addition, two features on the northwest ansa newly discovered by \citet{boccaletti18} need to be accommodated---this might be done by shifting the avalanche zone to the northwest (potentially improving the fit to velocities observed to the southeast) and experimenting with variable avalanche histories to reproduce the relative clump photometry (whose uncertainties seem large). 
    Other problems---the extreme stellar mass loss rates, the evacuation of micron-sized grains in potential violation of the disk's blue color (\citealt{fitzgerald07}; \citealt{schuppler15}), and underestimation of clump velocities---might also be ameliorated by situating the avalanche zone closer to the star, where the stellar wind blows more strongly and where disk material persists (\citealt{macgregor13}; \citealt{matthews15}; this paper; see also \citet{thebault&kraj18} for the advantages in launching avalanche seed material from an inner belt).
    Lorentz forces from the wind might also help radial acceleration.}}
Our lower limit of $\sim \SI{400}{km}$ on the perturbing body size {is consistent} with the $\sim \SI{400}{km}$ radius for their Varuna-like progenitor (for reference, Varuna is among the larger objects in the Kuiper belt).  
Moreover, because in their scenario it takes $\sim \SI{3e4}{yr}$ to clear the debris, which is only a small fraction of the AU Mic system age of $\sim \SI{20}{Myr}$, we should expect many such progenitors, to ensure an order-unity probability for observing the aftermath of a single catastrophic collision today.  
We might expect a Varuna-sized object to be disrupted every $\sim \SI{3e4}{yr}$, for a total of $\sim 700$ such progenitors over the system age---this is a minimum estimate, as there could exist many more such progenitors that could take longer than the system age to be destroyed.  
An ensemble of $\sim 10^3$ \SI{400}{km} bodies with density \SI{2}{g.cm^{-3}} amounts to a total mass of $\sim \SI{0.09}{M_\earth}$, safely below the upper limit on the total disk mass of ${\sim \SI{1.8}{M_\earth}}$ established by the present work. Also relevant to the results presented here is the work of \citet{matra19}, who perform a similar analysis connecting vertical structure to the mass of stirring bodies in the $\beta$ Pic system.

\subsection{Stellar Activity}
\label{subsection: stellar activity}

Our analysis indicates with high confidence that AU Mic exhibited long-term millimeter variability over the $\sim \SI{1.5}{yr}$ observation period, with stellar
flux densities preferred by our modeling varying by more than a factor of two (\SI[parse-numbers=false]{160^{+20}_{-30}}{\mu Jy} to \SI{390 \pm 20}{\mu Jy}) from one observation to the next.
Our values are comparable to the best-fit \SI{1.28}{mm} flux density of \SI{320 \pm 60}{\mu Jy} from \citet{macgregor13}, which is equivalent to \SI{290 \pm 50}{\mu Jy} when scaled to our observing wavelength of \SI{1.35}{mm} assuming a $\lambda^{-2}$ scaling for thermal emission.
It seems likely that this variability is distinct from the \SI{6}{minute} flare spanning two orders of magnitude that occurred during the June observations. 
That being said, we cannot rule out the possibility that the observations caught the star in varying states of flare decay, especially as \citet{white96} note that shorter-wavelength radio flares tend to have slower decay timescales.

{Further evidence for the long-term variability of AU Mic is provided by \citet{alekseev&kozhevnikova17}, who compile from the literature 16 epochs of $V$-band observations spanning more than two decades.
As shown in Figure 3 of \citet{alekseev&kozhevnikova17}, the star exhibits variability on the order of 0.3 magnitudes both within individual epochs and across timespans of several years.
\cite{macgregor16} also detect variability on scales of minutes to months in \SI{9}{mm} VLA observations of AU Mic, with typical flux densities $\sim \SI{650}{\mu Jy}$ on 2013 May 9, $\sim \SI{1000}{\mu Jy}$ on 2013 May 11, and $\sim \SI{650}{\mu Jy}$ on 2013 June 21.}
Simulations of radio emission from low-mass stars allow for variability of more than a factor of two over the entire longitudinal extent of a star \citep{llama18}; {\citet{rodono86} find AU Mic to exhibit $V$-band variability with amplitudes of up to 0.3 magnitudes and a period equivalent to the 4.85-day rotation period of the star \citep{kiraga&stepien07}.}
\citet{cox85} detect small-scale ($\sim 50\%$) variations over  a two-week period in \SIrange{2}{20}{cm} observations of AU Mic with the VLA, possibly due to 'several independent mini-flares' or rotational modulation of the star.
In fact, factor of $\sim 2$ variability over months to years is typical of 'quiescent' microwave emission from cool stars \citep{guedel94}.
Regardless of the origin of AU Mic's short- and long-term variability, we emphasize the importance of accounting for stellar emission/variability in observations of circumstellar disks around low-mass stars, especially in light of the Proxima Centauri flare discovered by \citet{macgregor18}.

\section{Conclusions}
\label{section: conclusion}

We have presented new \SI{1.3}{mm} ALMA observations of thermal dust emission from the debris disk around AU Mic at nearly two times finer angular resolution than previous observations. 
Both the vertical and radial structure of the disk are resolved.
MCMC analysis suggests that the radial structure exhibits an increasing surface density profile to $\sim \SI{41}{au}$ and is best characterized by an inner radius $r_{in} \sim \SI{23}{au}$ and power law exponent $p \sim 0.8$, although a lower-likelihood solution exists at $r_{in} \sim \SI{9}{au},\ p \sim 2.8$
We see no indication of a millimeter complement to the fast-moving features detected in the optical by \citet{boccaletti15}, but the data are suggestive of an additional ring of dust at $\sim \SI{10}{au}$.
Models with a ring interior to the main disk provide a better fit to the data, but the difference is not clearly statistically significant, possibly due to eccentricity or non-coplanarity of the residual feature.
MCMC analysis prefers an aspect ratio $h = 0.031$, corresponding to a vertical scale height $H \sim \SI{1.2}{au}$ in the outer regions of the disk.
Our analysis suggests that this is not an upper limit; a model with vertically resolved structure provides a statistically improved fit over a model with unresolved vertical structure at a $4 \sigma$ confidence level.
Furthermore, the disk vertical FWHM derived from parametric modeling corresponds well with image-domain estimates of the beam-subtracted FWHM of the emission perpendicular to the disk plane.

By comparing our measurement of the scale height to the steady-state collisional modeling of \citet{pan&schlichting12} we are able to place constraints on the mass and sizes of bodies stirring AU Mic's disk.
In the lower-limit case where collisional velocity damping is inefficient, the stirring bodies would have a radius of \SI[parse-numbers=false]{400^{+60}_{-50}}{km}, corresponding to a characteristic mass $\sim 20$ times less than that of Pluto.
On the other hand, velocity damping may balance stirring; this condition allows us to place an upper limit of $\sim \SI{1.8}{M_\earth}$ on the product of the square root of the number of stirring bodies and their individual masses.
This result, also an upper limit because it neglects the possibility of perturbations by external bodies, implies a maximum planet mass of $\sim \SI{1.8}{M_\earth}$ in the case of a single stirring planet. 
These results rule out the presence of a gas giant or Neptune analog embedded within the outer disk, but are suggestive of a significant population of asteroids at least \SI{400}{km} in size.
Such a population has been inferred on independent grounds using time-variable infrared observations \citep{chiang&fung17}.

Looking forward, the scale height measurement presented in this work could be combined with other measurements of AU Mic's scale height at widely-separated (sub)millimeter wavelengths.
This would allow the size-dependent velocity dispersion and internal strengths of bodies in AU Mic's collisional cascade to be constrained, testing the assumption of collisional cascade theory that velocity dispersion is constant with grain size.
Our measurements of the AU Mic system provide a proof of concept that spatially resolved observations of the vertical structure at millimeter wavelengths can constrain the presence of Uranus and Neptune analogs and even large Kuiper belt object analogues, which are undetectable by standard planet-detection techniques.  
Applying this technique to other high-inclination debris disks with a range of central stellar masses will provide unique constraints on the prevalence of large perturbing bodies throughout the Galaxy.

\section*{Acknowledgements}
C.D. is sponsored by a NASA CT Space Grant Undergraduate Research Fellowship and Wesleyan University's Research in the Sciences Fellowship.  
C.D., A.M.H., E.C., and K.F.~gratefully acknowledge support from NSF grant AST-1412647.  
M.P.~gratefully acknowledges support from NASA grants NNX15AM35G and NNX15AK23G.
H.E.S~gratefully acknowledges support from NASA grant NNX15AK23G.
E.I.C.~acknowledges support from the National Science Foundation.
J.M.C.~acknowledges support from the National Aeronautics and Space Administration under grant No. 15XRP15\_20140 issued through the Exoplanets Research Program.

This paper makes use of the following ALMA data: ADS/JAO.ALMA\#2012.1.00198.S.  
ALMA is a part\-nership of ESO (representing its member states), NSF (USA) and NINS (Japan), together with NRC (Canada), MOST and ASIAA (Taiwan), and KASI (Republic of Korea), in cooperation with the Republic of Chile.  
The Joint ALMA Observatory is operated by ESO, AUI/NRAO and NAOJ.  
The National Radio Astronomy Observatory is a facility of the National Science Foundation operated under cooperative agreement by Associated Universities, Inc.

This work has made use of data from the European Space Agency (ESA) mission {\it Gaia} (\url{https://www.cosmos.esa.int/gaia}), processed by the {\it Gaia}
Data Processing and Analysis Consortium (DPAC, \url{https://www.cosmos.esa.int/web/gaia/dpac/consortium}). 
Funding for the DPAC has been provided by national institutions, in particular the institutions participating in the {\it Gaia} Multilateral Agreement.

\software{
\texttt{CASA}  \citep{mcmullin07},
\texttt{MIRIAD} \citep{sault95},
\texttt{NumPy} \citep{van2011numpy}, 
\texttt{Astropy} \citep{astropy},  
\texttt{Pandas} \citep{mckinney}, 
\texttt{emcee} \citep{foreman-mackey13},
\texttt{Uncertainties}, \url{http://pythonhosted.org/uncertainties} 
}
\clearpage
\bibliography{AU_Mic_bibliography.bib}
\end{document}